\documentclass[twocolumn,trackchanges]{aastex631}
\usepackage{graphicx}
\usepackage{float}
\usepackage{amsmath}
\usepackage{color}
\usepackage{silence}
\newcommand{\Nc}{N$_{\mathrm{C}}$}
\newcommand{\aNc}{$\overline{N_{C}}$}
\newcommand{\afi}{$\overline{f_{i}}$}

\received{February 17, 2022}
\revised{March 30, 2022}
\accepted{April 10, 2022}

\shorttitle{Linking PAH Characteristics with Galaxy Properties}
\shortauthors{Maragkoudakis \emph{et al.}}
\graphicspath{{./}{figures/}}

\begin{document}

\title{Linking Characteristics of the Polycyclic Aromatic Hydrocarbon Population with Galaxy Properties: A Quantitative Approach Using the NASA Ames PAH IR Spectroscopic Database}

\correspondingauthor{A.~Maragkoudakis}
\email{Alexandros.Maragkoudakis@nasa.gov}

\author[0000-0003-2552-3871]{A.~Maragkoudakis}
\affiliation{NASA Ames Research Center, MS 245-6, Moffett Field, CA 94035-1000, USA}
\affiliation{Oak Ridge Associated Universities, Oak Ridge, TN, USA}

\author[0000-0002-4836-217X]{C.~Boersma}
\affiliation{NASA Ames Research Center, MS 245-6, Moffett Field, CA 94035-1000, USA}

\author[0000-0002-8341-342X]{P.~Temi}
\affiliation{NASA Ames Research Center, MS 245-6, Moffett Field, CA 94035-1000, USA}

\author[0000-0002-1440-5362]{J.D.~Bregman}
\affiliation{NASA Ames Research Center, MS 245-6, Moffett Field, CA 94035-1000, USA}

\author[0000-0002-6049-4079]{L.J.~Allamandola}
\affiliation{NASA Ames Research Center, MS 245-6, Moffett Field, CA 94035-1000, USA}

\begin{abstract}

    Utilizing the data and tools provided through the NASA Ames PAH IR Spectroscopic Database (PAHdb), we study the PAH component of over 900 \textit{Spitzer}-IRS galaxy spectra. Employing a database-fitting approach, the average PAH size, the PAH size distribution, and PAH ionization fraction are deduced. In turn, we examine their connection with the properties of the host galaxy. We found that PAH population within galaxies consists of middle-sized PAHs with an average number of carbon atoms of \aNc{} = 55, and a charge state distribution of $\sim$40\% ionized -- 60\% neutral. We describe a correlation between the 6.2/11.2 \micron{} PAH ratio with the ionization parameter ($\gamma\equiv(G_{0}/n_{\rm e})(T_{\rm gas} / 1\ \mathrm{K})^{0.5}$), a moderate correlation between the 8.6/11.2 \micron{} PAH ratio and specific star-formation rate, and a weak anti-correlation between $\gamma$ and M$_{*}$. From the PAHdb decomposition we provide estimates for the 3.3 \micron{} PAH band, not covered by \textit{Spitzer} observations, and establish a correlation between the 3.3/11.2 \micron{} PAH ratio with \Nc. We further deliver a library of mid-IR PAH template spectra parameterized on PAH size and ionization fraction, which can be used in galaxy spectral energy distribution fitting codes for the modeling of the mid-IR PAH emission component in galaxies.

\end{abstract}


\section{Introduction} \label{sec:introduction}

The mid-infrared (mid-IR) spectra of numerous astrophysical sources, such as planetary and reflection nebula, the interstellar medium (ISM), star-forming regions, and consequently of entire galaxies, are dominated by prominent emission features at 3.3, 6.2, 7.7, 8.6, 11.2, and 12.7~\micron, attributed to polycyclic aromatic hydrocarbons \citep[PAHs;][]{Leger1984, Allamandola1985}. Extensive observations from the \textit{Infrared Space Observatory} (ISO) and the \textit{Spitzer} Space Telescope have established the omnipresence of PAH emission in galaxies in both the local and high-redshift Universe \citep[e.g.,][]{Genzel1998, Rigopoulou1999, Armus2007, Gordon2008, ODowd2009, Riechers2014, Li2020}, as well as in diverse galactic environments, i.e., dominated by either star-formation, active galactic nuclei (AGN) activity, or a combination thereof \citep[e.g.,][]{Smith07b, Alonso-Herro2014, Esparza-Arredondo2018}.

Emission from PAHs has been extensively used as a tool in galactic and extra-galactic studies, as the spectral characteristics can be directly tied to the prevailing local astrophysical conditions of the emitting regions. This is because the spectrum is a reflection of the composition of the underlying PAH population, which, in turn, is determined by the astrophysical and astrochemical environment. For example, their abundance and emission properties depend on the metallicity and radiation field characteristics \citep[e.g.,][]{Galliano2008, Sandstrom2012}. However, PAHs do not only passively respond to their environment, but also drive many of its aspects, e.g., they regulate the charge balance through their high electron affinity \citep[e.g.,][]{BakesTielens1994}. Via a set of empirical relations, PAH emission is employed to infer the local physical conditions (i.e., the strength of the local UV field, $G_{\rm 0}$, electron density, $n_{\rm e}$, and gas temperature, $T_{\rm gas}$), in both resolved galactic sources \citep[e.g.,][]{Fleming2010, Rosenberg2011, Boersma2015} and galaxies \citep[e.g.,][]{Galliano2008}. Furthermore, for galaxies in particular, PAH luminosity has been calibrated and used as a tracer of the integrated and spatially resolved star-formation rate (SFR), particularly for metal-rich and dust-rich galactic environments \citep[e.g.][]{Calzetti2007, Shipley2016, Maragkoudakis2018b}.

PAH emission can be responsible for some 5-30\% of the total IR emission observed from galaxies \citep[][]{Helou2000, Smith07b}, which establishes their significance. In galaxy studies, the modeling of the spectral energy distribution (SED) from (spectro-)photometric observations has grown in popularity over the recent years as a method for determining global and spatially resolved galaxy properties \citep[][]{Conroy2013, Ciesla2014, Chevallard2016, Ciesla2018, Nersesian2019, Enia2020, Johnson2021}. Its applicability relies heavily on a proper modeling of the entire mid-IR spectrum as getting the energy balance correct plays a crucial role. That is, the energy absorbed by dust and PAHs in the UV-optical range directly translates to the energy emitted in the mid- and far-IR domain. Therefore, a robust framework and modeling of the three main ``dust'' components: PAHs; very small warm grains (VSGs); and big relatively cold grains (responsible for the emission beyond $\lambda \sim$100~\micron), are essential in galaxy studies. 

The implementation of the PAH emission component in dust emission models \citep[e.g.,][]{Draine2007, Draine2014, Dale2014} is considered not yet complete. The dust emission models of \cite{Draine2007}, for instance, are based on a dust mixture of amorphous silicate and graphite grains, and PAHs. These models separate dust emission into: \textit{(i)} diffuse emission from the general stellar dust population, and \textit{(ii)} emission from dust that is connected to star-forming regions, illuminated with a variable radiation field and parameterized with the fraction of PAHs locked-up in the total dust mass ($q_{\textrm{PAH}}$). 
These approaches typically don't take into account the effect of PAH structure and/or symmetry on the emerging spectrum, which can be significant \citep[e.g.,][]{Boersma2010, Andrews2015}. Furthermore, the high spectral resolution observations to be returned by the James Webb Space Telescope (\textit{JWST}) are expected to reveal unprecedented insight into the PAH (sub\nobreakdash-)populations and (sub\nobreakdash-)features, whose emission characteristics are highly-sensitive to molecular edge structure \citep[e.g.,][]{Peeters2017}.

Among the goals of this work is to supplement and enhance the PAH emission component of galaxy dust emission models by delivering a set of galaxy PAH emission spectral templates, derived from a large collection of PAH molecules of various sizes, molecular edge structures, and in different charge states. To achieve this, we performed a quantitative examination of the PAH characteristics in different galactic environments and their connection with fundamental galaxy properties (i.e., SFR, specific SFR (sSFR), stellar mass (M$_{\rm *}$), strength of the (local) radiation field ($G_{\rm 0}$), metallicity ($Z$)), utilizing the NASA Ames PAH IR Spectroscopic Database\footnote{\href{https://www.astrochemistry.org/pahdb/}{www.astrochemistry.org/pahdb/}} \citep[PAHdb hereafter;][]{Boersma2014a, Bauschlicher2018, Mattioda2020}. PAHdb data, models and software tools have already been successfully employed to probe the PAH properties in interstellar medium (ISM) sources \citep[][]{Cami2010, Boersma2013, Boersma2014a, Boersma2014b, Boersma2015, Boersma2018, Zang2019, Shannon2019}. Furthermore, we aim to establish whether PAH band strength ratios can be empirically calibrated into quantitative PAH ionization fractions ($f_{\rm i}$) and/or the PAH ionization parameter $\gamma$, the latter linking PAH ionization with $G_{\rm 0}$, $n_{\rm e}$, and $T_{\rm gas}$, as  has been successfully done for ISM sources. This work relies on the combined data from five \textit{Spitzer} Legacy Programs, culminating in over 900 spectra (see Section~\ref{sec:Sample}). The work described here, referred to as Paper~I, will be followed up by a paper (Paper~II) going into deeper detail on the sensitivity of the results to different model components, band strength measuring methods, pools of PAH molecules, and spectral wavelength range.

This paper is structured as follows: Section~\ref{sec:Sample} describes the sample of galaxies, its breakdown into different activity classes, and assesses the quality of the \textit{Spitzer}-IRS spectra. Section~\ref{sec:Specdecomp} provides the details of our analyses, including the spectral decomposition and modeling. The results and a discussion of their implications are presented in Section~\ref{sec:Results}. The paper is concluded in Section~\ref{sec:Summary} with a summary and conclusions.

\section{The Sample} \label{sec:Sample}

Section~\ref{subsec:Sample} describes how the sample was obtained and its make up. This is followed in Section~\ref{subsec:specquality} with an assessment of the quality of the spectra. Next, in Section~\ref{subsec:activityclass}, our method for assigning activity classes is described. 

\subsection{\textit{Spitzer} Legacy Programs} \label{subsec:Sample}
Our sample draws from the following five \textit{Spitzer} Legacy Programs, with data obtained directly from the he NASA/IPAC Infrared Science Archive\footnote{\href{http://irsa.ipac.caltech.edu/}{irsa.ipac.caltech.edu}} (IRSA):

\begin{enumerate}
    \item The Spitzer Infrared Nearby Galaxies Survey \citep[SINGS;][]{SINGS}. SINGS consists of present-day nearby galaxies ($d<30$ Mpc), primarily mapping their circumnuclear galactic regions.
    
    \item The Spitzer SDSS GALEX Spectroscopic Survey \citep[SSGSS;][]{ODowd2011}. SSGSS consists of normal, star-forming galaxies at $z<0.2$, color selected based on the 5.8 \micron{} surface brightness and 24~\micron\ flux. 
    
    \item The Spitzer SDSS Statistical Spectroscopic Survey \citep[S5;][]{Schiminovich2008}. S5 is an optically selected statistical sample of star-forming galaxies selected from the SDSSS at $z<0.1$. S5 is an extension of the SSGSS sample over a larger set of SDSS sources and narrower redshift range. 
    
    \item The 5 milli-Jansky Unbiased \textit{Spitzer} Extragalactic Survey \citep[5MUSES;][]{Wu2010}. 5MUSES is a 24 \micron{} flux-limited sample ($f_{24\, \micron}>5$ mJy) of intermediate redshift ($\overline{z} \sim 0.144$) designed to bridge the gap between the bright, nearby star-forming galaxies, and much fainter distant sources. 
    
    \item The Great Observatories All-Sky LIRG Survey \citep[GOALS;][]{Armus2009}. GOALS consists mostly of low-redshift ($z<0.09$) Luminous Infrared Galaxies (LIRGs), with a range of interaction stages (major mergers, minor mergers, and isolated galaxies). 
\end{enumerate}

The galaxies from these Legacy Programs cover a wide range in M$_{*}$, SFR, and $Z$ (Table~\ref{tab:sampleprops}) and sample various activity classes and galaxy environments, such as star-forming galaxies (SFGs), active galactic nuclei (AGN); which include Seyfert (Sy) and low-ionization nuclear emission-line regions (LINERs), composite systems of combined starburst and AGN contribution (CO), as well as LIRGs at various interaction stages (see Section~\ref{subsec:activityclass}). There is a large diversity among the sample in terms of physical environment, as well as distance--which translates to different sizes on the sky and thus differences in the spatially probed regions i.e., nuclear vs circumnuclear, resolved vs unresolved, etc. and observational parameters. In this work we do not focus on these differences, but rather examine the average PAH characteristics wholesale and their connection with general galaxy properties.

The galaxy properties were collected from the MPA-JHU catalog\footnote{\href{MPA-JHU DR7 release of spectrum measurements}{wwwmpa.mpa-garching.mpg.de/SDSS/DR7}} or directly from the literature, when available. Table~\ref{tab:machine} in Appendix~\ref{tab:machine} provides an overview of the collected properties per Legacy Programs. It is noted that these properties are not always derived using the same methods/calibrations (we refer the reader to the Legacy Programs themselves for details). 

The total number of galaxy spectra  considered is 910 and all cover both the SL (5-14.2~\micron) and LL (13.9-39.9~\micron) segments. In this work we make use of only the SL segment, which holds all the prominent PAH emission features, as well as as a large number of atomic fine-structure lines and rotational lines from molecular hydrogen. In Paper~II we will examine the impact of utilizing the LL segment in our analysis.

\begin{deluxetable*}{lccccc}
    \tablenum{1}
    \tablecaption{Summary of Galaxy Sample Properties per Legacy Program\label{tab:sampleprops}}
    \tablewidth{0pt}
    \tablehead{\colhead{Program} & \colhead{$N_{\rm gal}$\tablenotemark{a}} & \colhead{$z$} & \colhead{SFR} & \colhead{M$_{*}$} & \colhead{$Z$\tablenotemark{b}} \\
    \colhead{} & \colhead{} & \colhead{} & \colhead{[M$_{\odot}\,\mathrm{yr}^{-1}$]} & \colhead{[M$_{\odot}$]}}
    \startdata
    SINGS & 57 & $0.0006\le z \le 0.02 $ & $-1.5 < \log(\mathrm{SFR}) < 1.0$ & \ldots & $7.81 < Z < 8.6$ \\
    SSGSS & 94 & $0.03 \le z \le 0.2$ & $-0.9 \le \log(\mathrm{SFR}) \le 1.8$ & $9.2 \le \log(\mathrm{M}_{*}) \le 11.3$ & $8.5 \le Z \le 8.9$ \\
    S5 & 291 & $0.05 < z < 0.1$ & $-1.2 \le \log(\mathrm{SFR}) \le 1.6$ & $9.1 \le \log(\mathrm{M}_{*}) \le 11.5$ & $8.2 \le Z \le 8.8$ \\
    GOALS & 189 & $0.003 < z < 0.09$ & $1.2 \le \log(\mathrm{SFR}) \le 2.7$ & $10.3 \le \log(\mathrm{M}_{*}) \le 11.8$ & \ldots \\
    5MUSES & 279 & $0.02 < z < 0.2$ & $-0.9 \le \log(\mathrm{SFR}) \le 1.8$ & $9.4 \le \log(\mathrm{M}_{*}) \le 11.4$ & $8.4 \le Z \le 8.8$ \\
    \enddata
    \tablenotetext{a}{Total number of galaxies from program considered here.}
    \tablenotetext{b}{$12 + \log(\mathrm{O}/\mathrm{H})$.}
\end{deluxetable*}


\subsection{Quality Assessments} \label{subsec:specquality}

All SL spectra were moved to rest-frame wavelengths (\micron) and their signal put into units of flux density (Jy). The majority of spectra in the sample show the prominent PAH emission features. Though, there are a few that show very weak or no PAH emission at all. A wide range in dust continua is observed, from relatively weak to dominating the spectrum. In some rare cases there are apparent calibration issues, which predominantly affect the SL\emph{2} (5-7.4~\micron) part of the spectrum. Others have been affected by solar flares, causing a large spread in signal-to-noise ratio (S/N) \citep[SSGSS;][]{ODowd2011}. For the spectra and details on the data reduction we refer to the Legacy Programs themselves.

The quality of each spectrum was assessed using the S/N of the total spectrum (S/N$_{\rm tot}$) and the S/N of the 11.2~\micron\ PAH band (S/N$_{\rm 11.2}$). Specifically, we require: \textit{1.} S/N$_{\rm tot} > 2.4$ which is the value that corresponds to the fifth percentile with the median being 9.55, and \textit{2.} S/N$_{\rm 11.2} \ge 3$, in order to acquire spectra of resolved PAH features. The S/N$_{\rm tot}$ is computed as the ratio of the integral of a spectrum's intensity and uncertainty. S/N$_{\rm 11.2}$ is computed with respect to an underlying, straight-line continuum. The flux density (S) of the 11.2~\micron\ PAH band is simultaneously fitted with the sum of a 1$^{\rm st}$ order polynomial and a Gaussian line profile between 11.1-11.3~\micron. For the the noise (N) the uncertainties of the fit were propagated.


\subsection{Galaxy Activity Classification} \label{subsec:activityclass}

The SEDs of galaxies can show significant variation depending on activity class (e.g., SFGs, AGN), which is also true for their mid-IR spectra. For example, AGN systems show suppressed PAH intensities, which has been related to either PAH destruction \citep[e.g.,][]{Smith07b}, induced by the harsh radiation field associated with AGNs, or the PAH emission being swamped by the AGN continuum \citep[e.g.,][]{Armus2007}.

For those galaxies with optical spectroscopic coverage (516 galaxies), we derive their activity classifications following \cite{Maragkoudakis2018a}, which uses the combined classification acquired from the three optical emission line diagnostics ([O\,\textsc{iii}]$/$H$\beta$ -- [N\,\textsc{ii}]$/$H$\alpha$, [O\,\textsc{iii}]$/$H$\beta$ -- [S\,\textsc{ii}]$/$H$\alpha$, [O\,\textsc{iii}]$/$H$\beta$ -- [O\,\textsc{i}]$/$H$\alpha$; \citealt{BPT}; \citealt{Kewley2001}; \citealt{Kauffmann2003}; \citealt{Kewley2006}; \citealt{Schawinski2007}) to classify galaxies into SFGs, Sy, LINER, and CO. For the SSGSS, S5, and 5MUSES Legacy data, we used optical spectroscopic information from the MPA-JHU catalog, and for the SINGS galaxies the optical activity classification was obtained from \cite{Moustakas2010}. 

An alternative classification scheme using mid-IR spectra takes the equivalent width of the 6.2~\micron\ PAH band (EQW$_{\rm 6.2}$ \citep[][]{Armus2007} to classify galaxies as SFGs, AGN, and CO. Following \cite{Armus2007}, we defined star-forming dominated systems ($\equiv$SFGs) as those having a EQW$_{\rm 6.2}>1.5~$\micron, composite systems (both starburst and AGN contributions; $\equiv$CO) those with $0.2<\textrm{EQW}_{\rm 6.2}\le0.5$~\micron, and AGN dominated systems those with EQW$_{\rm 6.2}\le0.2$~\micron.
    
There is 86.3\% agreement between the optical and EQW$_{\rm 6.2}$ classification, where the Sy and LINERs were considered as AGNs. A typical case for divergence was the classification of Sy as CO when using the EQW$_{\rm 6.2}$ method. The EQW$_{\rm 6.2}$ method breaks the sample down in 84\% SFGs, 10\% AGN, and 6\% CO. 

\section{Analysis} \label{sec:Specdecomp}

The overall data analysis largely follows that described in, e.g., \cite{Boersma2018} and consists of the following five steps.

\begin{enumerate}
    \item \textit{Isolating the PAH Emission Spectrum--}Emission that originates in PAHs is separated from the underlying stellar and dust continuum, as well as from emission lines associated with molecular hydrogen and atomic species (Section~\ref{subsec:pahfit}). The spectra are simultaneously compensated for extinction and PAH band strengths are determined. 
    \item \textit{Fitting the PAH Emission Spectrum--}The 5.2-14.2~\micron\ SL PAH emission spectrum is fitted using PAHdb, which allows breaking down the emitting PAH family into contributing PAH sub-classes, i.e., charge, size, composition, and structure (Section~\ref{subsec:pahdb}).
    \item \textit{Estimating Uncertainties of PAHdb-derived Parameters--}A Monte Carlo technique is employed to estimate PAHdb uncertainties (Section~\ref{subsubsec:pahdbdecomp}).
    \item \textit{Extrapolating the PAH spectrum--}The fitted PAH spectrum is extrapolated to provide a complete 3-20~\micron\ PAH spectrum (Section~\ref{subsubsec:extrapolated}).
    \item \textit{Calibrating Qualitative PAH Proxies--}The connection between PAH band strength ratios measured from step 1 and the PAH subclass decomposition from step 2 is examined (Section~\ref{subsec:pahip}).
\end{enumerate}

\subsection{PAHFIT} \label{subsec:pahfit}

A 5.2-14.2~\micron\ spectral decomposition was performed with the Python version of \textsc{pahfit}\footnote{\href{https://github.com/PAHFIT/pahfit}{github.com/PAHFIT/pahfit}}. \textsc{pahfit} is an Astropy \citep{Astropy2018} affiliated package. The Python version of \textsc{pahfit} is based off the IDL version by \cite{Smith07b} and is currently being ported as part of a \textit{JWST} Early Release Science (ERS) Program\footnote{\href{https://www.stsci.edu/jwst/science-execution/approved-programs/dd-ers/program-1288}{\textit{JWST} ERS program ID: 1288}}. 
The reader is referred to \cite{Smith07b} for a description of \textsc{pahfit}. A comparison between the results from the IDL and Python versions of \textsc{pahfit} is done in Appendix~\ref{sec:IDLvsPy}.

The quality of the spectral decomposition and modeling is assessed in terms of the ratio between the integral of the absolute fit residuals and that of the input spectrum ($\sigma_{\textsc{pahfit}}$). A near-perfect fitted spectrum would result in $\sigma_{\textsc{pahfit}}\sim0$, with  poorer fits yielding values upwards. 

\begin{figure*}
    \includegraphics[width=\linewidth]{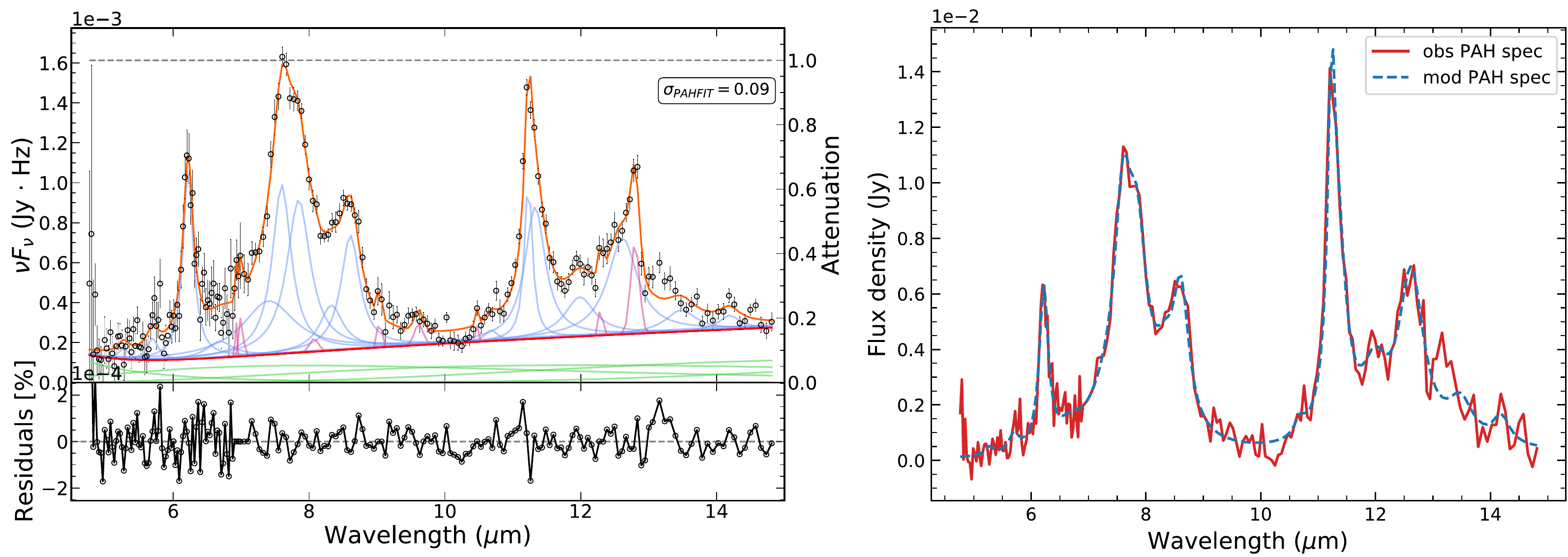}
        \caption{\textbf{Left:} \textsc{pahfit} decomposition of the 5.2-14.5~\micron\ \textit{Spitzer}-IRS spectrum of the galaxy SDSS J093001.33+390242.0 (ID: S5\_11\_4, in the S5 catalog). The fit (orange line) is synthesized using the following components: Dust features (light blue lines), atomic and H$_{\rm 2}$ lines (magenta lines), continuum (green lines; the total continuum emission is shown as a red line), and attenuation (dashed black line). $\sigma_{\textrm{PAHFIT}}$ is provided in the box (see Section~\ref{subsec:pahfit}) for details. \textbf{Right:} Modeled (blue dashed line) and observed (red line) 5.2-14.5~\micron\ isolated PAH spectrum (see Section~\ref{subsec:pahfit} for details) for the galaxy SDSS~J093001.33+390242.0.}
    \label{fig:pahfit}
\end{figure*}

The left panel of Figure~\ref{fig:pahfit} demonstrates the \textsc{pahfit} decomposition of the 4.2-14.2~\micron\ spectrum of SDSS~J093001.33+390242.0, a SFG from the S5 Legacy program, and has a $\sigma_{\textsc{pahfit}}$ of 0.09, the median value for the entire sample. The fit provides a good separation between the different components with 1-1.5\% residuals that are predominantly at the noise level. For this particular galaxy, the \textsc{pahfit} modeled extinction is negligible, while the silicate extinction at 9.7~\micron\ ($\tau_{9.7}$) has a median of 0.1 when considering the entire sample. In the current analysis, only sources with $\tau_{9.7} < 2$ were considered. This constraint was applied to exclude highly attenuated (U)LIRGs from the GOALS sample, for which the \textsc{pahfit} decomposition, although producing a low $\sigma_{\textsc{pahfit}}$, may not be optimal given the difficulty of applying MIR dust models normally used for star-forming and starburst galaxies to heavily obscured LIRGS \citep[see e.g.,][]{Dopita2011}. Additional \textsc{pahfit} decomposition examples for various galaxy classes with diverse properties are provided in Appendix~\ref{sec:additional_examples} (Figure~\ref{fig:pahfit_examples}).

The \textsc{pahfit} results are used to construct two isolated PAH spectra: \textit{(i)} The \textsc{pahfit} modeled PAH spectrum from combining the different Drude components, and \textit{(ii)} The observed PAH spectrum from subtracting the dust continuum, stellar continuum, atomic and H$_{2}$ line \textsc{pahfit} components from the observed spectrum, and accounting for extinction. It is noted that both isolated PAH spectra are not independent, as the stellar, dust continuum, and Gaussian components are also used to construct the observed PAH spectrum. However, the observed PAH spectrum retains both emission not matched by any \textsc{pahfit} component and the noise.  The right panel of Figure~\ref{fig:pahfit} compares the two isolated PAH spectra for the galaxy SDSS~J093001.33+390242.0.

The figure shows both spectra possessing the salient PAH features, with the differences the residual shown in the left panel of Figure~\ref{fig:pahfit}, which are dominated by the noise. However, there is some subtle variances of note. For example just red of the 12.7~\micron\ PAH band, excess emission is present that resembles a distinct band that is not matched by \textsc{pahfit}. Also, some of the galaxy spectra show an 11.2~\micron\ feature that has emission at its blue wing that is not matched by \textsc{pahfit} and could be linked to the 11.0~\micron\ satellite feature. Our main analysis relies on the observed (\texttt{obs}) PAH spectrum, but the appendices provide an analogue analysis for the modeled (\texttt{mod}) one.


PAH band strengths, along with the atomic fine-structure and rotational line strengths, are  determined from \textsc{pahfit}. Here, combinations of individual and blended Drude profiles are used to derive strengths for the 6.2, ``7.7'', 8.6, 11.2, 12.7, 13.5, and 14.2~\micron\ PAH bands, analogue to the IDL version of \textsc{pahfit}. The PAH band strengths determined this way have been made available online\footnote{\href{https://www.astrochemistry.org/pah\_galaxy\_properties/}{www.astrochemistry.org/pah\_galaxy\_properties/}}. In Paper~II we will compare and contrast this approach with others for recovering PAH band strength, e.g., the spline method \citep[e.g.,][]{Uchida2000, Hony2001, Peeters2002, Galliano2008, Boersma2014a, Peeters2017, Xie2018}. 


\subsection{PAHdb} \label{subsec:pahdb}

The isolated PAH emission spectra obtained from Section~\ref{subsec:pahfit} are further analyzed using PAHdb. Specifically, the galaxy PAH spectra are fit utilizing the AmesPAHdbPythonSuite\footnote{\href{https://github.com/PAHdb/AmesPAHdbPythonSuite}{github.com/PAHdb/AmesPAHdbPythonSuite}}, which is based off the IDL suite\footnote{\href{https://github.com/PAHdb/AmesPAHdbIDLSuite}{github.com/PAHdb/AmesPAHdbIDLSuite}} of tools, along with version 3.20 of PAHdb's library of density functional theory (DFT) computed spectra. This library holds 4233 quantum-chemically calculated absorption spectra of PAHs with various structures, charge states, sizes, and compositions. This allows the analysis of astronomical spectra without adopting any ad hoc assumptions on the characteristics and state of the underlying PAH population \citep[e.g.,][]{Li2001}, as all the properties of the individual PAH molecules used to synthesize the modeled spectrum can be fully recovered.

\subsubsection{The Pool of Astronomically Relevant PAH Spectra in PAHdb} \label{subsubsec:pahsample}

A pool of spectra from PAH molecules that meet the astronomically relevant criteria as laid out in \cite{Bauschlicher2018} is considered. Specifically, PAHs containing more than 20 carbon atoms (\Nc), which are assumed to survive the harsh interstellar environments \citep[e.g.][]{Allamandola1989, Puget1989}. Furthermore, either only spectra from ``pure'' PAHs or those containing nitrogen and no aliphatic side groups are used. 
Except for fullerenes, the spectra from fully dehydrogenated PAHs are also not considered.
The pool thus arrived at 2968\footnote{The pool can be obtained on the PAHdb \href{https://www.astrochemistry.org/pahdb/}{website} with the query string \texttt{"magnesium=0 oxygen=0 iron=0 silicium=0 chx=0 ch2=0 c>20 h>0"}, further adding in the fullerenes, with UIDs: 717, 720, 723, 735, 736, and 737. Note the returned pool will be dependent on the version of the library used, here 3.20.} spectra. 

\subsubsection{Model Parameters} \label{subsubsec:pahdbmodel}

In order to model the astronomical PAH \emph{emission}, the DFT-computed \emph{absorption} data \citep[see e.g.,][]{Mattioda2020} needs to be converted into \emph{emission} spectra. This requires taking into account: \textit{(i)} the radiation field that PAHs are exposed to, \textit{(ii)} the molecular relaxation process after excitation, \textit{(iii)} the line profile and the  width of the emitting bands, and \textit{(iv)} possible band shifts due to anharmonic effects.

Here, a range of excitation energies is considered; 6, 8, 10, and 12~eV, together with an emission model that takes the entire emission cascade into account \citep[see e.g.,][]{Boersma2011, Boersma2013}. Gaussian line profiles are used with a FWHM of 15~cm$^{\rm -1}$. 
A 15~cm$^{\rm -1}$ redshift is applied to mimic (some) anharmonicity effects, a typical value adopted in the literature \citep[e.g.,][]{Bauschlicher2009}. The impact of using different line profiles or omitting a redshift \citep[][]{Mackie2018} will be explored in Paper~II. While the different excitation energies will help discern any sensitivity to the radiation field, they also accommodate construction of the galaxy PAH emission template spectra (Section~\ref{subsec:templates}). Our main analysis relies on PAH spectra computed using an excitation of 8~eV.

\subsubsection{Fitting, Uncertainties, and Breakdown} \label{subsubsec:pahdbdecomp}

\begin{figure*}
    \includegraphics[width=\linewidth]{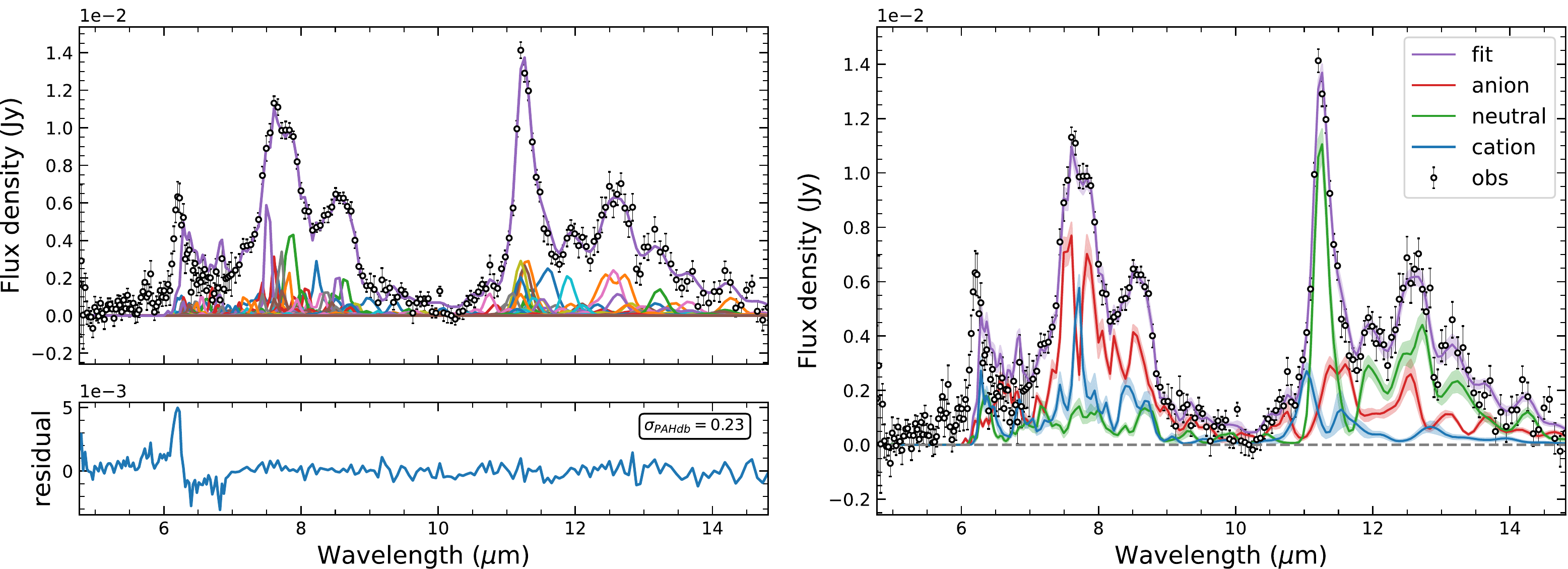}
    \caption{PAHdb-fit and charge decomposition of the 5.2-14.5\micron\ PAH spectrum of the galaxy SDSS~J093001.33+390242 (circles). \textbf{Left:} Contribution from each of the 46 individual PAH molecules (colored lines), along with the residual spectrum (bottom panel). The inset shows $\sigma_{\rm PAHdb}$ (see Section~\ref{subsec:specquality}). \textbf{Right:} Breakdown in terms of charge. The uncertainties obtained from the Monte Carlo sampling are shown for each component as a shaded envelope. See Section~\ref{subsubsec:pahdbdecomp} for details.}
    \label{fig:pahdbdecomp}
\end{figure*}

The PAHdb fitting was performed using a Non-Negative Least Chi-square minimization approach \citep[NNLC;][]{Desesquelles2009}. Figure~\ref{fig:pahdbdecomp} (left panel) presents the results of the PAHdb-fit of the 5.2-14.5~\micron\ PAH spectrum of the galaxy SDSS~J093001.33+390242 (same galaxy as in Figure~\ref{fig:pahfit}). The figure shows overall a good fit. Additional PAHdb fitting and breakdown examples for various galaxy classes with diverse properties are provided in Appendix~\ref{sec:additional_examples} (Figure~\ref{fig:pahdb_examples}).

Similarly as for \textsc{pahfit}, PAHdb uncertainties ($\sigma_{\rm PAHdb}$) are quantified as the ratio between the integrals of the absolute residuals over that of the input spectrum (see also Section~\ref{subsec:pahfit}). While $\sigma_{\textsc{pahfit}}$ shows a value of 0.09, $\sigma_{\rm PAHdb}$ is considerably larger at 0.23; though a value that is on par with those reported elsewhere for similar fits \citep[e.g., ][]{Bauschlicher2018, Ricca2021}. 

It has been put forward that this relatively large ``error'' is driven by PAHdb-fits having a systematic difficulty in matching certain wavelength regimes \citep[e.g.,][]{Bauschlicher2018}. For example, as is apparent in Figure~\ref{fig:pahdbdecomp}, the blue side of the 6.2~\micron\ PAH band is not well matched. This particular difficulty has been attributed to the limited number of, and variation in PANHs in PAHdb's libraries. PANHs are, to date, the only PAHs that are able to accommodate the very blue 6.2~\micron\ emission \citep[e.g.,][]{Peeters2002, Hudgins2005, Ricca2021}. Furthermore, the emission from $\sim$5-6~\micron\ is attributed to PAH overtone and combination bands \citep[e.g.,][]{Boersma2009}, which are not part of the PAH emission spectra modeled here. In Appendix~\ref{sec:PAHdbPAH62} we quantify the fraction of the 6.2~\micron\ PAH band recovered by PAHdb with respect to that by \textsc{pahfit}.

To gain a better understanding of these systematics and for a fairer quality criterion, we turn to computing $\sigma_{PAHdb}$ for wavelength ranges that bracket the main PAH features. For SDSS~J093001.33+390242 these are 1.02, 0.57, 0.04, 0.04, 0.05, 0.1, and 0.1 for 
$\sigma_{5.2}$, $\sigma_{6.2}$, $\sigma_{7.7}$, $\sigma_{8.6}$, $\sigma_{11.2}$, $\sigma_{12.0}$, and $\sigma_{12.7}$, respectively. These values indicate that indeed much of the ``error'' is systematic in nature. The median $\sigma_{\rm, PAHdb}$ for the entire sample is 0.3, while that for $\sigma_{PAHdb,11.2}$ is 0.07, a value on par with that found for $\sigma_{\textsc{pahfit}}$.


Although the identification of individual astronomical PAH molecules in the mid-IR is (practically) impossible due to band overlap, PAHdb-fits do allow probing the general state of the PAH population in terms of subclasses. It has been shown that the PAH charge and size composition can be determined with a high degree of reliability \citep[e.g.,][]{Andrews2015}. In turn, they can be tied to the PAH ionization fraction and PAH size distribution.

To obtain uncertainties for the derived  parameters, we turn to a Monte Carlo (MC) approach. Here, the observed PAH spectra are perturbed 1000 times within their uncertainties and fitted. Subsequently, for each galaxy the statistical average and the standard deviation are determined.

The right panel of Figure~\ref{fig:pahdbdecomp} shows the PAHdb-fitted 5.2-14.5~\micron\ PAH spectrum of SDSS~J093001.33+390242 again, but now broken down in terms of charge. The figure shows a small amount of variation for the fit itself, with somewhat more for the three charge components (anion, neutral, and cation) individually. 

From the PAHdb-fits we construct a PAH size distribution. Here, we take a PAH's effective radius (in \AA) as $a_{\rm eff}=\sqrt{A/pi}$, where $A$ is the area of the PAH computed by multiplying the sum of the number of rings by the area of a single (3-8-member) ring. Subsequently, a power-law is fitted and the power-law index ($\alpha$) is recorded. Besides $\alpha$, we also record the average PAH size in terms of number of carbon atoms ($overline{N_{\rm C}}$), the two-charge state PAH ionization fraction ($f_{\rm i} \equiv n_{\mathrm{PAH^{+}}}/(n_{\mathrm{PAH^{0}}}+n_{\mathrm{PAH^{+}}})$), and the different hydrogen adjacency classes; all with their associated Monte Carlo determined uncertainties.

\subsubsection{Extrapolating the Fitted PAH Spectrum} \label{subsubsec:extrapolated}

\textit{JWST} is anticipated to provide a wealth of spectroscopic information from 0.6-28.8~\micron. Consequently, acquiring information that spans this entire range is valuable for gaining early insights as well as benchmarking \textit{JWST} observations. Therefore, we extrapolated the PAHdb-fitted PAH spectra to cover the 3-20~\micron\ range.

Furthermore, information gained on the 3.3~\micron\ PAH band would be of value for non-high redshift \textit{Spitzer} observations \citep[][]{Sajina2009}, where the band is absent. 3.3~\micron\ PAH emission intensity is particularly sensitive to PAH size. When ratioed to the 11.2~\micron\ PAH band strength, it provides a robust tracer of PAH size \citep[][]{Allamandola1989, Schutte1993, Mori2012, Ricca2012, Croiset2016, Maragkoudakis2018b, Maragkoudakis2020}. In addition, the 3.3~\micron\ PAH band can put tight constraints on the total energy put into PAHs in dust models \citep[e.g.,][]{Lai2020}. Moreover, extrapolated spectral data beyond 15~\micron\ would be of use for those \textit{Spitzer}-IRS observations that lack LL data. 

\begin{figure}
    \includegraphics[width=\linewidth]{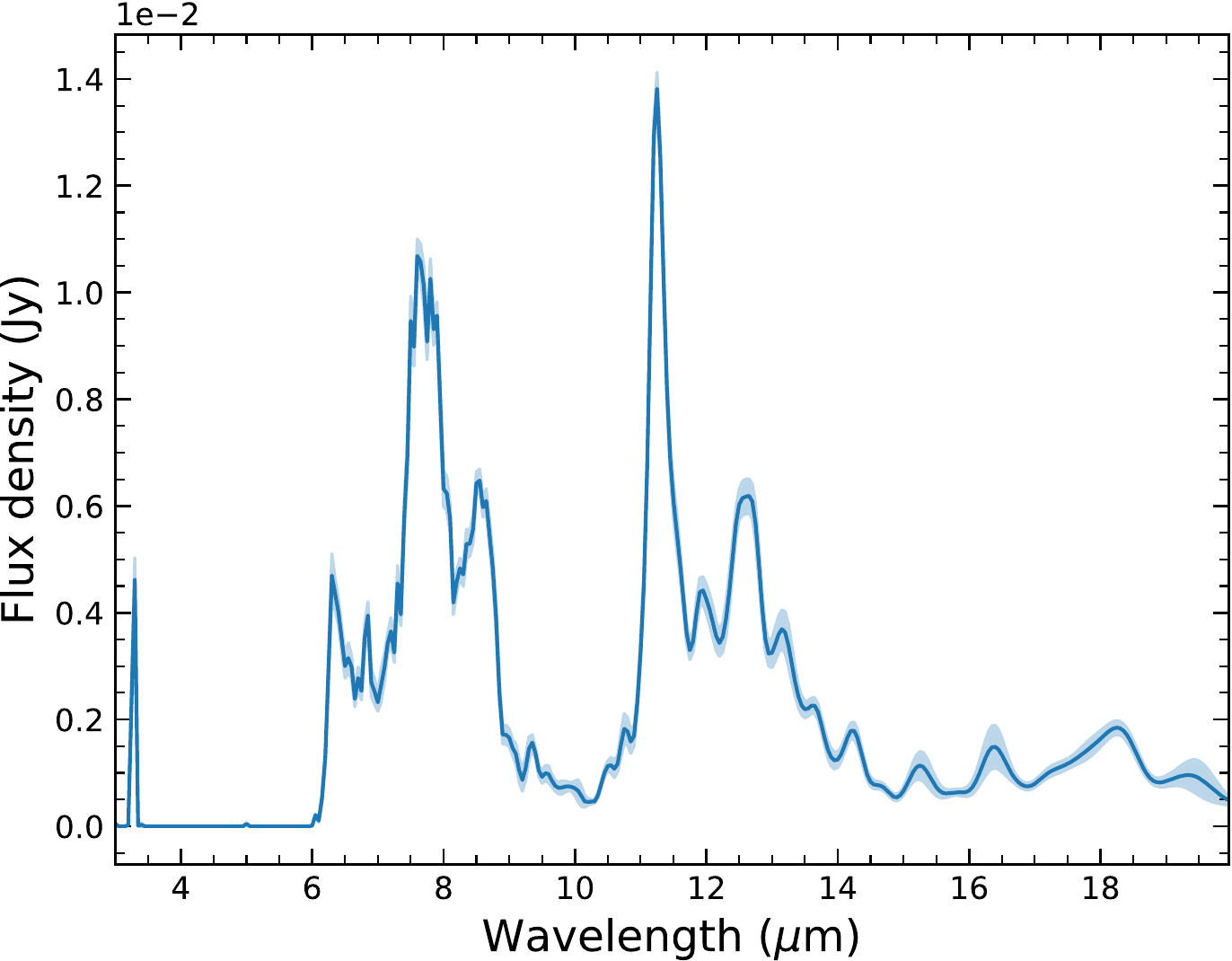}
    \caption{The extrapolated 3-20~\micron\ PAH spectrum (blue line) of the galaxy SDSS~J093001.33+390242 with the associated uncertainty shown as the shaded envelope. See Section~\ref{subsubsec:extrapolated} for details. \label{fig:predictedspec}}
\end{figure}

The extrapolated spectra are synthesized by co-adding the 3-20~\micron\ spectra of each PAH contributing to a fit, where each PAH spectrum has been multiplied by its fitted weight. Since this is done as part of the Monte Carlo sampling, the final extrapolated spectrum is accompanied by uncertainties. Figure~\ref{fig:predictedspec} presents the extrapolated 3-20~\micron\ spectrum of the galaxy SDSS~J093001.33+390242. The figure reveals a distinct, simple 3.3~\micron\ PAH band, as well as substantial substructure longwards of 15~\micron. Note that the astronomical 3.3~\micron\ feature is due to highly energetic C-H stretching modes that couple overtone and combination bands and is bracketed with more structure than the harmonic band shows in Figure~\ref{fig:predictedspec} \citep[][]{Mackie2018, Maltseva2018}. Thus, the actual structure of any astronomical 3.3~\micron\ PAH feature will differ substantially from that shown in the figure. The strength of the extrapolated 3.3~\micron\ PAH band is determined by fitting a Gaussian.

\subsection{Calibrating Qualitative PAH Proxies} \label{subsec:pahip}

PAH band strength ratios are sensitive to, and therefore reflect, astrophysical conditions. For example, the coupled C--C stretching and C--H in-plane bending modes increase considerably upon ionization compared to C--H stretching modes \citep[e.g.,][]{Allamandola1999}, and thus the 6.2/11.2 and 7.7/11.2~\micron\ PAH band strength ratios are frequently used to measure the PAH ionization balance, and subsequently--via empirical calibrations--the ionization parameter $\gamma$ ($\equiv (G_{0}/n_{\rm e})(T_{\rm gas} / 1\ \mathrm{K})^{0.5}$) \citep[e.g.,][]{BregmanTemi2005, Galliano2008}. 

The PAHdb decomposition quantifies the cation and neutral PAH contribution, which allows for a \emph{quantitative} calibration of the empirical, \emph{qualitative} proxies used to probe variations in the properties of the emitting PAH populations, such as charge, size, or structure. For example, the PAH charge balance is often approximated as: 

\begin{equation} \label{eq:npah_62_112}
    \frac{n_{\mathrm{PAH^{+}}}}{n_{\mathrm{PAH^{0}}}} \propto \frac{I_{6.2}}{I_{11.2}}\ ,
\end{equation}
\citep[see][]{Boersma2014a, Boersma2016, Boersma2018} where the PAH cation and neutral densities ($n_{\mathrm{PAH^{+}}}$ and $n_{\mathrm{PAH^{0}}}$, respectively) are obtained from the PAHdb decomposition. When assuming two accessible ionization states and parameters applicable for circumcoronene\footnote{Note that there is some uncertainty associated with the parameters used that depends on the adopted electron recombination rates, where the difference between the classical and measured rates on very small PAHs can be nearly as high as an order of magnitude \citep[see][]{Tielens2005}.} (C$_{54}$H$_{12}$), which is considered representative of an average interstellar PAH (\Nc=50-100; \citealt{Croiset2016}), the PAH ionization parameter can be expressed as:

\begin{equation} \label{eq:npah_ioniz_param}
    \gamma \equiv(G_{0}/n_{\rm e})(T_{\rm gas}/1\ \mathrm{K})^{0.5}= 2.66\, (\frac{n_{\mathrm{PAH^{+}}}}{n_{\mathrm{PAH^{0}}}})\ [\times10^{4}\ \mathrm{cm}^{3}].
\end{equation}


Similarly, the correlation between the 11.2/3.3~\micron\ PAH band strength ratio and \Nc\ has been well-documented in the literature \citep[e.g.,][]{Ricca2012, Croiset2016, Maragkoudakis2020}

\section{Results and Discussion} \label{sec:Results}

We now investigate the distributions of the derived PAH properties (Section~\ref{sec:pahdistributions}), correlations among the derived PAH properties, and with galaxy properties, broken down per galaxy activity class (Section~\ref{subsec:PAH_Gal_correlate}). PAH band strength ratio calibrations are examined in Section~\ref{subsec:pahratios}. A library of template galaxy PAH emission spectra, quantified based on $\overline{N_{C}}$ and $\overline{f_{\rm i}}$, are presented in Section~\ref{subsec:templates}. Finally, we discuss some of the limitations associated with our methodology in Section~\ref{subsec:limitations}. 

During our analysis a number of parameters were established to assess the quality of the data and the reliability of the fitting results. We combined these now in such a way to establish three quality classes: Q1-Q3. All three classes meet the spectral quality requirements that S/N$_{\rm (tot)} \ge 2.4$ and S/N$_{\rm (11.2)} \ge 3$ (Section~\ref{subsec:specquality}). Each subsequent class sets less stringent constraints on the \textsc{pahfit} uncertainty ($\sigma_{\textsc{pahfit}}$; Section~\ref{subsec:pahfit}), and the PAHdb uncertainty ($\sigma_{\rm PAHdb,11.2}$; Section~\ref{subsubsec:pahdbdecomp}). For $\sigma_{\textsc{pahfit}}$ and $\sigma_{\rm PAHdb,11.2}$ the limits are set based on where their cumulative distributions hit 68, 95, and 99.7\%. This sets $\sigma_{\textsc{pahfit}}\le$ 0.11, 0.27 and 0.38, and $\sigma_{\rm PAHdb,11.2}\le $ 0.07, 0.12, and 0.26, for Q1, Q2 and Q3, respectively. Q1 has 386, Q2 698, and Q3 778 galaxy spectra. These quality classes provide different levels of confidence and can help interpreting the results. Our main results are based on Q2 data. A comparison between results derived using the different quality classes is provided in Appendix~\ref{sec:classcomp}. All data have been made available online\footnote{\href{https://www.astrochemistry.org/pah\_galaxy\_properties/}{www.astrochemistry.org/pah\_galaxy\_properties/}} (see Appendix~\ref{sec:mastertable}). 


\subsection{PAH Property Distributions} \label{sec:pahdistributions}

\begin{figure*}
    \includegraphics[width=\linewidth]{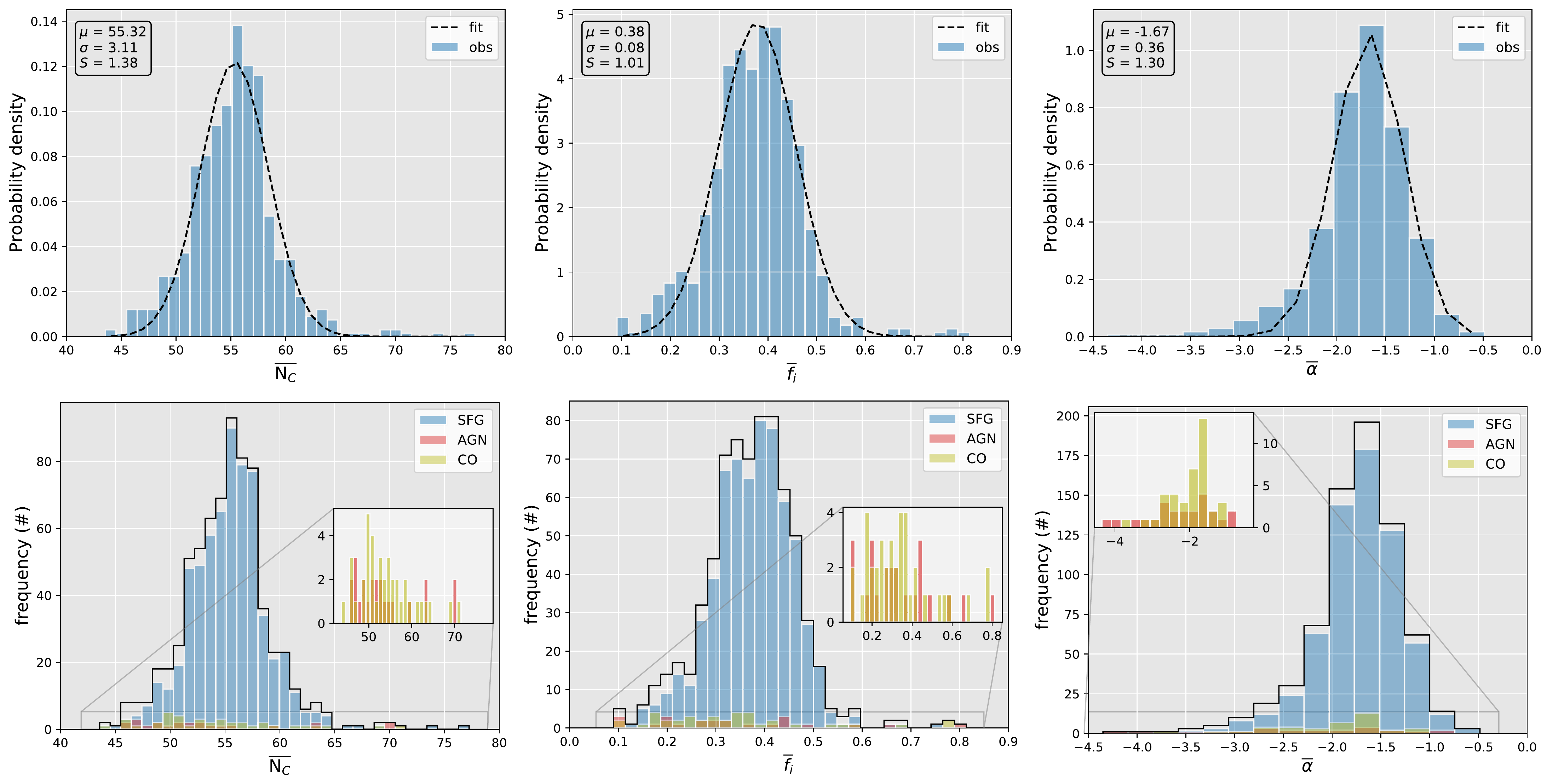}
    \caption{Distributions for the average number of carbon atoms ($\overline{N}_{\rm C}$; left panels), average ionization fraction ($\overline{f_{\rm i}}$; middle panels), and average power-law index of the PAH size distribution ($\overline{\alpha}$; right panels). \textbf{Top row}: Probability density distributions fitted with a Gaussian model (black dashed lines). The statistical mean ($\mu$), standard deviation ($\sigma$), and skewness (\textit{S}) are given in the boxes. \textbf{Bottom row}: Probability density distributions broken down by galaxy activity class (SFGs, AGN, CO). That of the combined classes are shown in black.}
    \label{fig:alldistributions}
\end{figure*}

From the PAHdb MC sampling analysis of each galaxy we collected a set of PAH properties, including \Nc, $f_{\rm i}$, PAH composition (``pure'' vs nitrogen containing), and hydrogen adjacency classes information (solo, duo, trio, quartet). A weighted average is computed using the inverse MC-determined uncertainties squared as weights. Figure~\ref{fig:alldistributions} presents the distributions for $\overline{N}_{\rm C}$, $\overline{f_{\rm i}}$, and of the power-law index $\overline{\alpha}$ from the size distribution fits. The number of bins and bin widths for each histogram is determined using Knuth's rule \citep[][]{Knuth2006}, which utilizes a Bayesian model and computes the posterior probability of the number of bins for a given data set without making any \emph{a priori} assumptions on the data. Each distribution was also fitted with a Gaussian model for comparison. Besides this, the statistical mean, standard deviation, and skewness were also determined. The results are summarized, alongside the parameters derived from the Gaussian distributions, in Table~\ref{tab:pahdistributions}.

\begin{deluxetable}{cccc}
    \tablenum{2}
    \tablecaption{PAH Property Distribution Attributes \label{tab:pahdistributions}}
    \tablewidth{0pt}
    \tablehead{\colhead{Parameter} & \colhead{$\mu \pm \sigma$} & \colhead{$\mu_{\rm fit} \pm \sigma_{\rm fit}$\tablenotemark{a}} & \colhead{\textit{S}\tablenotemark{b}}}
    \startdata
    $\overline{N_{C}}$ & $55.24 \pm 3.90$ & $55.32 \pm 3.10$ & 1.39 \\
    $\overline{f_{i}}$  & $0.37 \pm 0.09$ & $0.38 \pm 0.08$ & 1.01 \\
    $\overline{\alpha}$ & $-1.75 \pm 0.46$ & $-1.67 \pm 0.36$ & 1.31 \\
    \hline
    \colhead{Class} & \colhead{$\overline{N_{C}}$ ($\mu\pm\sigma$)} & \colhead{$\overline{f_{i}}$ ($\mu\pm\sigma$)} & \colhead{$\overline{\alpha}$ ($\mu\pm\sigma$)} \\
    \hline
    SFGs & $55.38 \pm 3.55$ & $ 0.38 \pm 0.08$ & $-1.71 \pm 0.41$ \\ 
    AGN & $53.75 \pm 7.08$ & $ 0.34 \pm 0.18$ & $-2.22 \pm 0.92$ \\
    CO & $53.87 \pm 5.93 $ & $ 0.32 \pm 0.15$ & $-2.00 \pm 0.56$ \\
    \enddata
    \tablenotetext{a}{Mean ($\mu_{\rm fit}$) and standard deviation ($\sigma_{fit}$) from Gaussian fit.}
    \tablenotetext{b}{Skewness ($S$)}
    \tablecomments{Statistical mean ($\mu$) and standard deviation ($\sigma$) of the PAH property distributions for all the galaxies (top) and their respective activity classes (bottom).}
\end{deluxetable}


The statistical mean and standard deviation for $\overline{N}_{C}$, \afi, and $\overline{\alpha}$ distributions are in good agreement with those obtained from the Gaussian modeling, indicating a normal distribution. The weighted average number of carbon atoms is \aNc = 55, with a range of $44 \le \overline{N_{C}} \le 77$. This average supports the choice of cicumcoronene in Section~\ref{subsec:pahip} and provides confidence for the inferred ionization fraction. The weighted average ionization fraction has a value \afi\ = 0.37, with a range of $0.09 \le$ \afi\ $\le 0.81$. For the mean power-law index of the PAH size distribution a value of $\overline{\alpha} = -1.75$ and a range of $-4.35 \le \overline{\alpha} \le -0.48$ is found.

\aNc\ is for all galaxies (taking the standard deviation into account) close to the characteristic value that separates small from large astrophysical PAHs \citep[\Nc$\simeq$50; e.g.,][]{Tielens2008}. Considering the diversity and mixture of galactic environments, even within a given galaxy, it is, on the one hand, expected that in regions with high radiation field intensity (or hardness) such as star-forming regions or the central regions of AGN hosts, \Nc\ will be higher as smaller PAHs are more effectively destroyed. On the other hand, studies on the Small Magellanic Cloud (SMC) \cite{Sandstrom2010, Sandstrom2012} suggest that in low-metallicity environments PAHs only form inside dense molecular clouds, and are therefore typically smaller, or that larger PAHs form less efficiently. Consequently in both scenarios small PAHs are formed that are more susceptible to destruction. Our derived value for \aNc\ indicates that the average PAH population in galaxies consists of middle-sized PAHs. Figuring out whether this is the result of smoothing between regions of low and high \Nc\ PAH populations or something else will require spatially resolved studies within galactic environments. 

For \afi\ we find a value of 0.37, indicating an average distribution of $\sim$40\% ionized -- 60\% neutral of the PAH population within galaxies. In regions with not much ionizing radiation (e.g., inter-arm diffuse ISM regions) PAHs would be mostly neutral, whereas in regions with high levels of ionization (e.g., PDRs associated with star-forming regions) PAHs are expected to be mostly ionized. The electron recombination rate depends on electron density, which is known to vary on both galaxy-wide and H\,\textsc{ii}-region scales \citep[e.g.,][]{Herrera-Camus2016, Kewley2019}. As such, the mostly, average, neutral PAH population found for our sample is likely the result of averaging the ionization fraction of a number of regions with different physical conditions. 

The average PAH size distribution is weighted towards smaller PAHs, with $\overline{\alpha}\simeq-1.75$. Often PAHs are considered the extension of grain size distribution into the molecular domain. Compared to the MRN \citep[][; $\alpha=-3.5$]{Mathis1977} grain size distribution
that of PAHs is less heavily weighted towards smaller PAHs. One possible explanation for this could be that dust grains can grow continuously, whereas PAHs only in discrete steps, i.e., one ring at a time.

Breaking down the the distributions for \aNc, \afi, and $\overline{\alpha}$ per activity class (SFG, AGN, and CO) provides no new insights, though this could change when more sources, evenly spread among the different classes, are added.\\

Recently, \cite{Silva-Ribeiro2022} used PAHdb to study the main type of PAH molecules and the local physical conditions of their irradiating sources in 12 galaxies, where the entire PAHdb library was employed. They concluded that PAH species containing 10-82 carbon atoms were the most abundant in their sample and the PAH population can contain up to 95\% of small species and 79\% of neutral PAHs, with molecules such as C$_{52}$H$_{18}$, C$_{10}$H$_{9}$N, and C$_{14}$H$_{11}$N being present in more than 75 per cent of their sample while removal of these molecules showed a worsening in the fit. However, inclusion of the entire PAHdb library of molecules can bias the results. In Section~\ref{subsubsec:pahsample} we stressed the importance of considering a pool of PAH(db) molecules that meet astronomically relevant criteria, such as the consideration of only molecules containing more than 20-30 carbon atoms, which can survive destruction via photo-dissociation.



\subsection{PAH Band Strength and PAHdb-derived Correlations} \label{subsec:pahratios}

\begin{deluxetable}{cccc}
    \tablenum{3}
    \tablecaption{Correlation and fit parameters\label{tab:correlations}}
    \tablewidth{0pt}
    \tablehead{\colhead{Relation} & \colhead{r$_{xy}$} & \colhead{$\alpha$} & \colhead{$\beta$}}
    \startdata
	6.2/11.2 -- $\gamma$ & 0.65 & 0.46 $\pm$ 0.06 & 0.23 $\pm$ 0.10\\
	7.7/11.2 -- $\gamma$ & 0.53 & 1.29 $\pm$ 0.23 & 1.46 $\pm$ 0.36\\
	8.6/11.2 -- $\gamma$ & 0.38 & 0.15 $\pm$ 0.03 & 0.48 $\pm$ 0.05\\
	3.3/11.2 -- $\overline{N_{C}}$ & -0.42 & -0.04 $\pm$ 0.01 & 3.22 $\pm$ 0.37\\
	$\gamma$ -- M$_{*}$ & -0.36 & -0.39 $\pm$ 0.06 & 5.57 $\pm$ 0.59 \\
	6.2/11.2 -- sSFR & 0.49 & 0.16 $\pm$ 0.01 & 2.25 $\pm$ 0.13\\
	7.7/11.2 -- sSFR & 0.40 & 0.99 $\pm$ 0.08 & 13.15 $\pm$ 0.78\\
	8.6/11.2 -- sSFR & 0.39 & 0.24 $\pm$ 0.03 & 3.17 $\pm$ 0.28\\
	12.0/11.2 -- duo/solo & 0.34 & 0.20 $\pm$ 0.02 & 0.002 $\pm$ 0.025 \\
	6.2/7.7 -- $\overline{N_{C}}$ & -0.17 & -0.005 $\pm$ 0.001 & 0.553 $\pm$ 0.078 \\
    \enddata
    \tablecomments{Pearson's correlation coefficient ($r_{xy}$), slope ($\alpha$), and intercept ($\beta$) values of the linear fits for the various relations between PAH band strength ratios, PAHdb-derived parameters, and galaxy parameters.}
\end{deluxetable}

\begin{figure*}
    \centering
    \includegraphics[width=\linewidth]{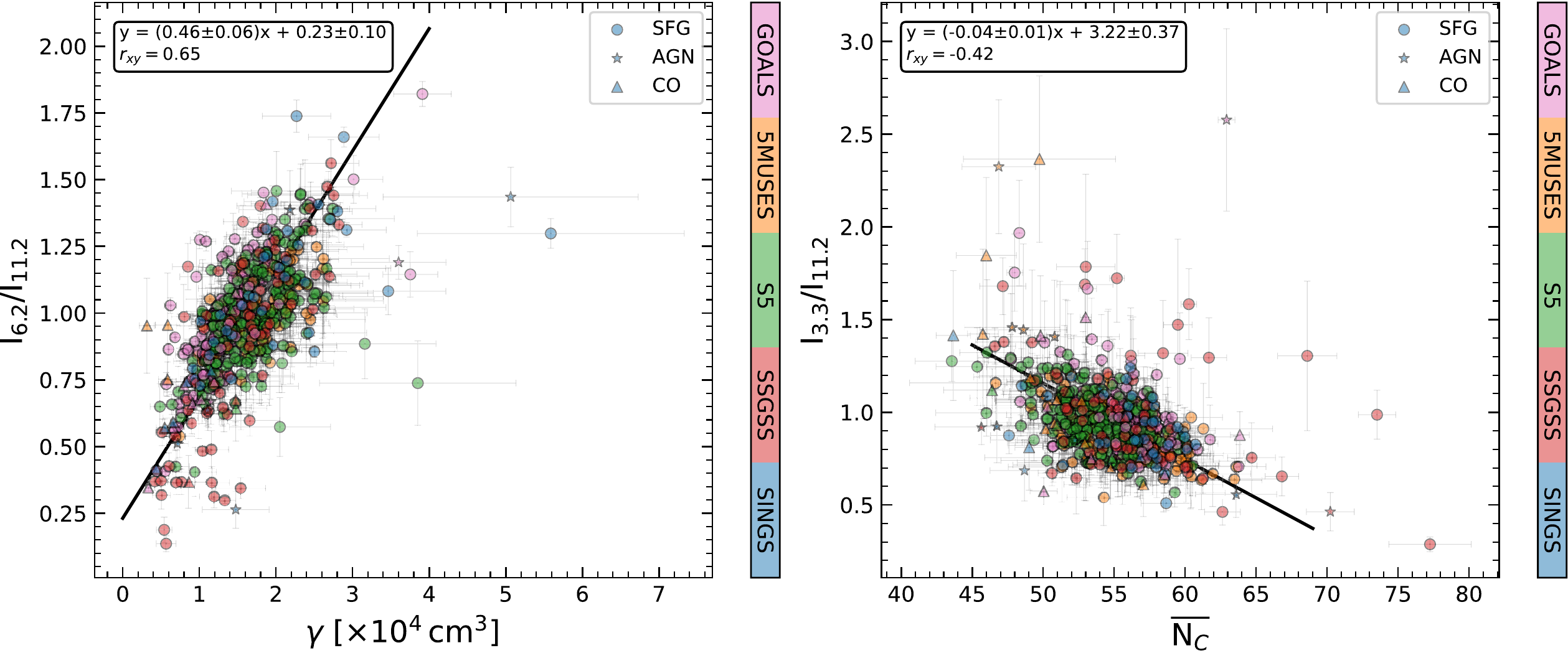}
    \caption{\textbf{Left:} The 6.2/11.2~\micron\ PAH band strength ratio versus the PAH ionization parameter ($\equiv\gamma$; Eq.~\eqref{eq:npah_ioniz_param}). \textbf{Right:} The 3.3/11.2~\micron\ PAH band strength ratio versus \aNc. SFGs are shown as circles, AGN as stars, and COs as triangles. Data points are color-coded based on their associated Legacy program. The linear fits are shown in black and the fit equation are given in the box, together with their Pearson's correlation coefficient ($r_{xy}$).}
    \label{fig:pahint}
\end{figure*}

Figure~\ref{fig:pahint} presents the correlations between the PAHdb-derived PAH ionization parameter and PAH size measures versus their well-known PAH band strength ratio proxies, i.e. the 6.2/11.2~\micron\ and 3.3/11.2~\micron\ PAH band strengths respectively. Although the 7.7/11.2 and 8.6/11.2~\micron\ PAH band strength ratios, which also probe PAH ionization, show a correlation with $\gamma$ (Appendix \ref{sec:weaker}, Figure \ref{fig:secondary}), the 6.2/11.2 -- $\gamma$ correlation is the tightest (Table \ref{tab:correlations}). The correlation is also evident when examining the respective samples, with the SSGSS and GOALS samples showing even tighter correlations with Pearson's correlation coefficient of $r_{xy}=0.79$ and $r_{xy}=0.75$ respectively. This relation allows for the estimation of the average ionization parameter within galaxies, directly from observationally measured quantities. Recovering these correlations for both PAH size and charge puts confidence in the overall PAHdb decomposition \citep[][]{Boersma2013}.

\begin{figure}
    \includegraphics[width=\linewidth]{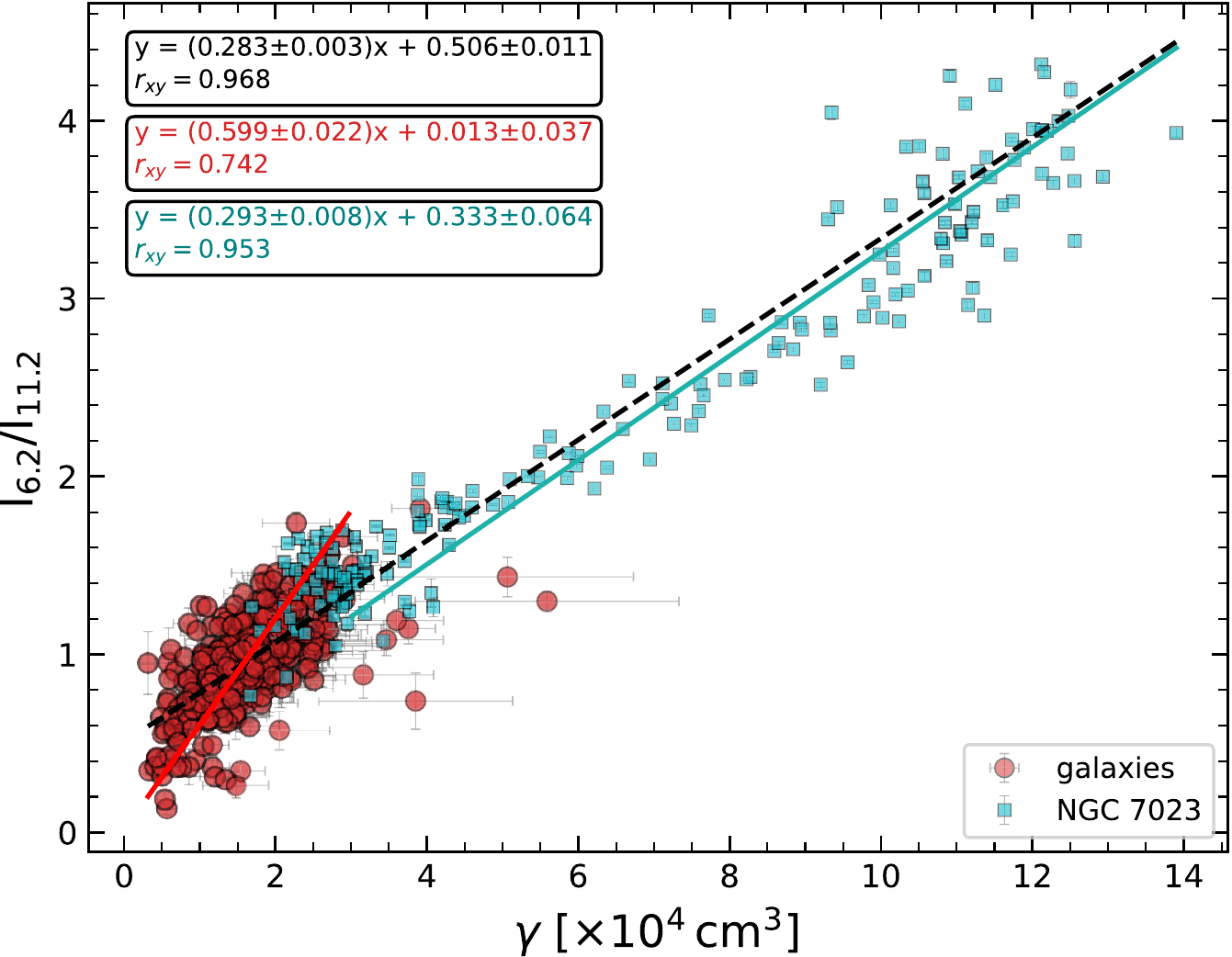}
    \caption{The 6.2/11.2~\micron\ PAH band strength ratio versus the PAH ionization parameter ($\equiv\gamma$; Eq.~\eqref{eq:npah_ioniz_param}), for galaxies (red circles) and the reflection nebula NGC~7023 (cyan squares). Linear fits are shown for the galaxy and NGC~7023 data (black dashed line), the galaxy and NGC~7023 data up to $\gamma = 3\times10^{4}\, \mathrm{cm}^{3}$ (red line), and NGC~7023 data alone (cyan line). The equations are shown with corresponding colors in the boxes, together with their Pearson's correlation coefficient ($r_{xy}$).}
    \label{fig:62_112_gamma_galaxies-NGC7023}
\end{figure}

Figure~\ref{fig:62_112_gamma_galaxies-NGC7023} extends the correlation presented in  Figure~\ref{fig:pahint} for the 6.2/11.2~\micron\ PAH band strength ratio versus $\gamma$ by including data on the reflection nebula NGC~7023 from \cite{Boersma2018}. The figure shows, what appears to be, a continuous transition from the galaxy to NGC~7023 data. To explore this further, three linear fits were performed: \textit{(i)} for the galaxy and NGC~7023 data combined, \textit{(ii)} the galaxy and NGC~7023 data combined up to $\gamma=3\times10^{4}$ cm$^{3}$, and \textit{(iii)} for the NGC~7023 data alone. For cases \textit{(i)} and \textit{(iii)} the fitted lines are very comparable, whilst differ with case \textit{(ii)}. 

Per \cite{Tielens2005}, the $\gamma$s for the galaxies fall between that of the general ISM and PDRs. Compared to the correlation found in (\citealt{Boersma2018}; their Figure~13), the galaxies overlap with the data on M17. Though M17 is classified as an H~\textsc{ii}-region, the data mostly reflects the shielded, more benign region behind the PDR. Indeed, this ISM-like environment might be more representative of the \emph{average} one encountered in star-forming galaxies, the bulk of galaxies considered here, as the PDRs associated with H~\textsc{ii}-regions are typically only a sliver on the sky compared to the molecular cloud that houses them. 

The correlation of the 3.3/11.2~\micron\ PAH band strength with \Nc\ is demonstrated for the galaxy samples (Figure~\ref{fig:pahint}, right panel) confirming the effectiveness of the 3.3/11.2~\micron\ PAH band strength ratio as a PAH size indicator in galaxies \citep[][]{Maragkoudakis2020}, and further demonstrating the efficiency of extrapolating the PAHdb-fitted spectra to include the 3.3~\micron\ band. For the 5MUSES and S5 samples the correlation is even tighter with $r_{xy}=-0.59$ and $r_{xy}=-0.55$ respectively. No correlation was found between the 6.2/7.7~\micron\ PAH band strength ratio--previously employed for PAH size determination \citep[e.g.,][]{Draine2001}, with \Nc\ ($r_{xy}=-017$, Table \ref{tab:correlations}; Appendix \ref{sec:weaker}, Figure \ref{fig:secondary}).


A rather weak trend is observed for the 12.0/11.2~\micron\ PAH band strength ratio vs the PAHdb-determined duo-to-solo hydrogen adjacency ratio (Appendix \ref{sec:weaker}; Figure \ref{fig:secondary}, and Table \ref{tab:correlations}). The 11.2 and 12.0~\micron\ PAH bands are typically associated with CH out-of-plane bending  modes of solo and duo hydrogens, respectively. Therefore, the 12.0/11.2~\micron\ PAH band strength ratio is often employed as a probe for PAH geometry, as solo modes are associated with long straight molecular edges, whereas duo and trio modes correspond to corners \citep[][]{Hony2001, Tielens2008}. All pairwise relationships between PAH band strength ratios with PAHdb-derived (and galaxy) properties are presented in Appendix \ref{sec:pahratio_pairplots} (Figures \ref{fig:PAH_ratios-galaxy_pairplots} and \ref{fig:PAH_ratios-PAHdb_pairplots}). 

\subsection{PAH and Galaxy Properties} \label{subsec:PAH_Gal_correlate}

\begin{figure*}
    \includegraphics[width=\linewidth]{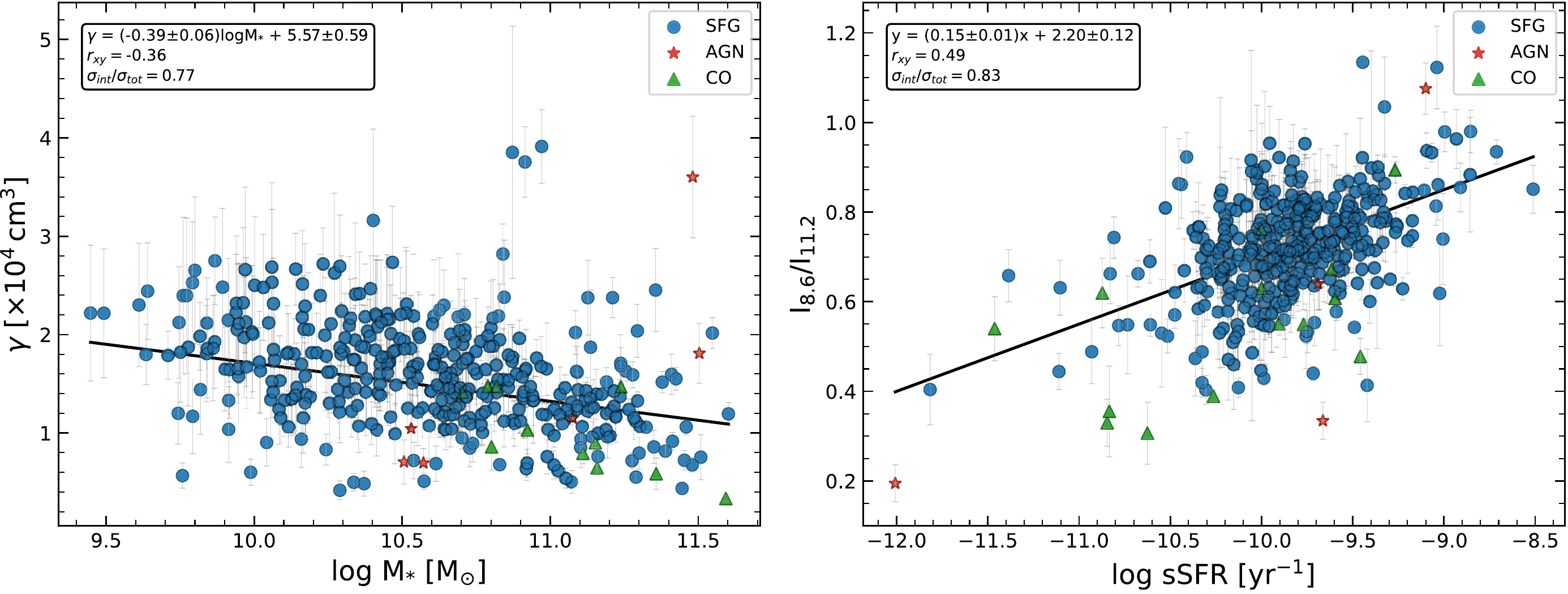}
    \caption{\textbf{Left:} The PAH ionization parameter $\gamma$ (Eq.~\eqref{eq:npah_ioniz_param}; Section~\ref{subsec:pahip}) vs stellar mass. \textbf{Right:} The 8.6/11.2\micron\ PAH band strength ratio vs sSFR. SFGs are shown as circles, AGN as stars, and CO as triangles. The black line shows the linear fit when assuming the presence of intrinsic scatter. The found fit equation is given in the box, together with the Pearson's correlation coefficient ($r_{xy}$), and the ratio of the intrinsic to the total scatter ($\sigma_{int}/\sigma_{tot}$).}
    \label{fig:PAH_gal}
\end{figure*}

We examined possible correlations between the PAHdb derived PAH properties and galaxy properties. Though weak (r$_{\rm xy}$=-0.36), one possible (anti-)correlation was found between M$_{*}$ and $\gamma$, which is presented in Figure~\ref{fig:PAH_gal} (left panel). Assuming such a correlation exists, we introduce an intrinsic scattering term per \cite{Williams2010} and find that it accounts for 77\% ($\sigma_{int}/\sigma_{tot}$) of the total scatter.
A possible origin for the large intrinsic scatter is the large variation in how the observations were acquired between the different Legacy programs, e.g., fiber or slit-spectroscopy, whether only nuclear- or circum-nuclear material or the galaxy as a whole was observed. 
The presence of such a correlation would support the view that in low-mass and low-metallicity galaxies the radiation field of young massive stars can more efficiently ionize PAHs due to the lower amounts of opacity and attenuating material (dust grains, gas and metals), whereas in more massive and metal-rich systems of higher column density material, the UV radiation of young stars is subjected more to attenuation, resulting in lower PAH ionization fractions. This (anti-)correlation becomes tighter when examining the Q1 quality class case (r$_{xy}=0.53$; see Appendix~\ref{sec:classcomp}).

A moderate correlation is identified between the PAH band strength ratio proxies of the PAH ionization (i.e., I$_{6.2}$/I$_{11.2}$, I$_{7.7}$/I$_{11.2}$, I$_{8.6}$/I$_{11.2}$) and the specific SFR (sSFR $\equiv$ SFR/M$_{*}$), with I$_{8.6}$/I$_{11.2}$ -- sSFR (Figure \ref{fig:PAH_gal}, right panel) being the tightest (r$_{xy}=0.49$; Table \ref{tab:correlations}), with the intrinsic scatter accounting for the 83\% of the total scatter. While the sSFR describes the relative SFR per galaxy stellar mass--and thus per atomic hydrogen (H\,\textsc{i}) mass content \citep[e.g.,][]{Catinella2010, Maddox2015}, the inverse of sSFR defines a timescale for the formation of the stellar population of a galaxy, where lower sSFR corresponds to older stellar populations for a constant or single-burst star-formation history \citep[][]{Whitaker2017}. In this sense, the correlation between the PAH ionization proxies with sSFR yields the increase of PAH ionization in systems of younger stellar populations and more recent star-formation episodes.

\begin{figure*}
    \includegraphics[width=\linewidth]{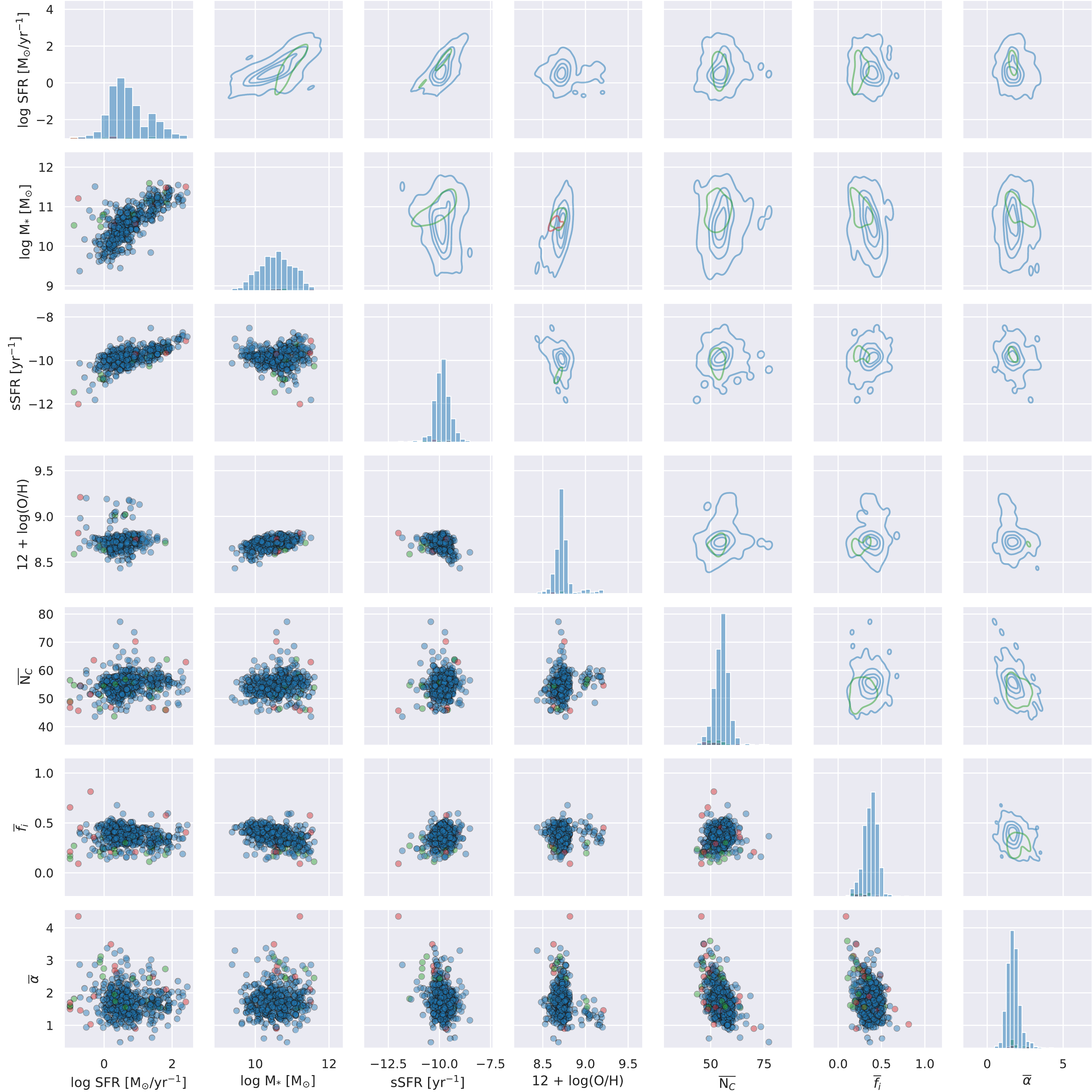}
    \caption{Pairwise relationships of galaxy (SFR, M$_{*}$, sSFR, 12+log(O/H)) and PAH (\aNc, \afi, $\overline{\alpha}$) properties (points), along with their respective distributions (histograms), and kernel density estimations (contours at 5 levels, corresponding to iso-proportions of the probability mass density, i.e., contours at 80, 60, 40, and 20\%). Colors correspond to galaxy activity classes, with SFGs shown in blue, AGN in red, and CO in green. \label{fig:pairplots}}
\end{figure*}

Figure~\ref{fig:pairplots} presents a pairwise comparison between galaxy and PAHdb-derived parameters. A similar comparison between PAH band strength ratios and galaxy parameters is given in Appendix~\ref{sec:pahratio_pairplots}. The figure shows that the well-documented relations like that between SFR and M$_{*}$ \citep[i.e., the galaxy main sequence,][]{Brinchmann2004, Elbaz2007, Whitaker2012, Maragkoudakis2017, Sanchez2020, Ellison2021} and the M$_{*}$-metallicity relation \citep[MZ-relation; e.g.,][]{Tremonti2004} are present. However, no \emph{obvious} correlation is found between the galaxy and PAHdb-derived parameters, other than the tentative one described between $\gamma$ and M$_{*}$, and similarly that of the PAH band strength ratios with sSFR (Figure~\ref{fig:PAH_ratios-galaxy_pairplots}).

PAH luminosity has been successfully calibrated as a tracer of SFR in galaxies \citep[e.g.][]{Calzetti2007, Shipley2016, Maragkoudakis2018b} and is commonly used as such. The cause for no correlation between the SFR and any of the PAH band strength ratios or PAHdb-derived parameters could be twofold: \textit{1.} The derived SFR is highly sensitive to the choice of calibrator, as different wavelengths tend to trace different stellar populations, different galaxy components and environments, and each calibrator has its own systematic effects and biases \citep[e.g.,][]{Calzetti2010, Murphy2011, Cluver2017, Mahajan2019, Kouroumpatzakis2021, Xie2022}; \textit{2.} The make up of the PAH population between galaxies is largely homogeneous \citep[][]{Andrews2015}.

To put these results on a stronger footing, a selection of (multiple) homogeneous sub-samples, broken down and grouped based on their physical conditions and environments (e.g., morphology, dust content, etc), probing similar regions (e.g., distance, size), with properties established using similar methods (calibrations) would be required. 

\subsection{Library of Template PAH Emission Spectra for Galaxies} \label{subsec:templates}

\begin{figure*}
    \includegraphics[width=\linewidth]{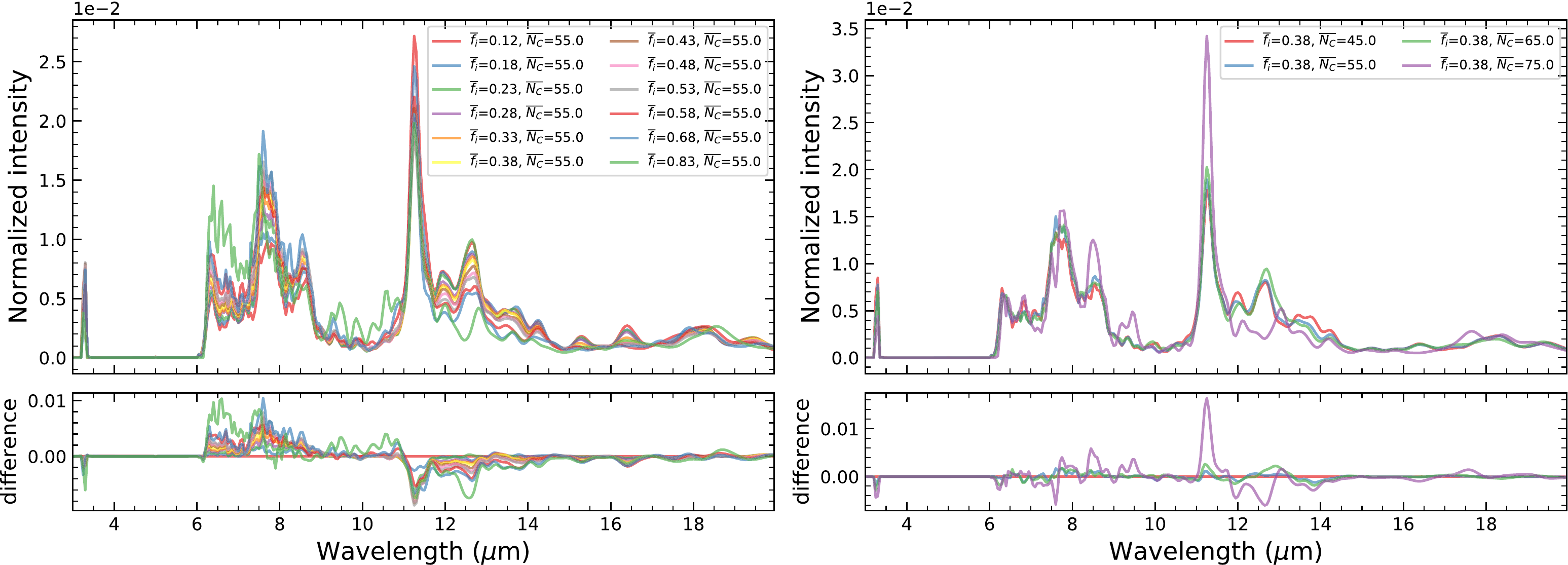}
    \caption{PAHdb-derived PAH emission templates. \textbf{Left:} Templates at fixed $\overline{N_{C}}$ and varying $\overline{f_{i}}$. \textbf{Right:} Templates at fixed $\overline{f_{i}}$ and varying $\overline{N_{C}}$. The spectra have been normalized (in $F_{\nu}$ units) on the total integrated flux. The bottom panels show the spectral differences between templates by subtracting each spectrum from the first spectrum in the set.}
    \label{fig:templates}
\end{figure*}

We have created a library of PAH emission spectra that can be used as templates in, e.g., galaxy SED modeling. The templates are parameterized on \aNc\ and $\overline{f_{\rm i}}$ 
Figure~\ref{fig:templates} compares a number of template spectra at fixed \aNc\ and varying $\overline{f_{\rm i}}$ (left), and vice versa (right). At fixed \aNc\, the PAH features in the 6-9~\micron\ wavelength range increase with increasing $\overline{f_{\rm i}}$, whereas the 3.3~\micron\ PAH band and those between 11-13~\micron\ decrease. At fixed $\overline{f_{\rm i}}$, the 3.3~\micron\ PAH band and those between 11-13~\micron\ increase with increasing \aNc, with no substantial variation between 6-9~\micron. The templates are offered for different radiation fields (6, 8, 10, and 12~eV) and have been made electronically available online\footnote{\href{https://www.astrochemistry.org/pahdb/templates_gal/}{www.astrochemistry.org/pahdb/templates\_gal/}}. 

\subsection{Final Thoughts} \label{subsec:limitations}

The results presented in this work depend strongly on the models, algorithms, and tools used to decompose the galaxy spectra, as well as the spectral resolution and quality of the astronomical spectra themselves.

With respect to \textsc{pahfit}, the resulting isolated PAH emission spectra and, subsequently, the measured band strengths are dependent on the adopted extinction model (i.e., fully mixed or a foreground screen of dust), as well as the shape of the extinction curve, which can vary in between galaxies \citep[][]{Salim2020}. \cite{Boersma2018} showed that the choice of extinction curve can have a significant impact on the measured PAH band strength ratios for H~\textsc{ii} regions. Specifically, they noted that \textsc{pahfit}-like fits appear sensitive to the shape of the used extinction curve through its interplay with other components of the spectra. That work showed that analyzing Spitzer spectral cubes using the R$_{\rm V}$=5.5 extinction curve from \cite{Weingartner2001} results in visual extinction ($A_{\rm V}$) maps with large pixel-to-pixel variations and discontinuities. However, when moving to the extinction curve from \cite{Chiar2006} a smooth continuous map is obtained. Subsequently, the choice of extinction curve impacts the derived PAH band strengths as well, which can be significant (up to $\sim$40\%). In addition, \textsc{pahfit} has been trained on star-forming and starburst galaxies and the application to heavily obscured galaxies here did not yield reliable results. 

PAHdb holds continuously growing libraries of quantum-chemically computed and laboratory measured PAH spectra and evolving fitting tools. The results presented here are based on version 3.20 of PAHdb's library of DFT-computed harmonic spectra. Regarding library content, it is important to understand its incompleteness. Efforts are underway to address these, specifically in matching the 6.2~\micron\ and the detailed structure of the 10-15~\micron\ region. Furthermore, the inclusion of anharmonicity in the computation of PAH emission spectra will help resolve any ambiguity with applying a 15~cm$^{\rm -1}$ red shift and the need for PAH anions to accommodate the extended red wings of the PAH bands \citep[e.g.,][]{BregmanTemi2005, Bauschlicher2009}.

In addition to the considerations described above and the current completeness state of PAHdb, putting forward a list of distinct dominant PAH molecules responsible for the galaxy PAH spectra would be premature, and as such it is avoided here. A comprehensive examination and quantification of degeneracies for robust conclusions on the dominant PAH populations in galaxies will be explored in Paper~II, along with the sensitivity of the results to PAH emission parameter choices.

The analysis and results presented in this work predominantly encapsulate the behavior of SFGs, as they make up a considerable fraction of the galaxies in the Legacy Programs.
Nonetheless, the current representation of the AGN and CO classes in the Legacy programs show similar distributions for the properties of their PAH populations to that of SFGs (Section~\ref{sec:pahdistributions}). 

\section{Summary and Conclusions} \label{sec:Summary}

We performed a detailed analysis of the PAH component of over 900 \textit{Spitzer}-IRS spectra from galaxies with different nuclear activity classes, utilizing the data, models and tools provided through PAHdb. The main conclusions from our work are summarized below.

\textit{(i)} The PAH population within galaxies consists of middle-sized PAHs with an average number of carbon atoms of \aNc{} = 55 and doesn't show much variation across the different activity classes (SFGs, AGN, CO). The found average two level PAH ionization fraction is $\overline{f_{i}}=0.37$,  $\overline{f_{i}}_{(\mathrm{SFG})}=0.38$, $\overline{f_{i}}_{(\mathrm{AGN})}=0.33$, and $\overline{f_{i}}_{(\mathrm{CO})}=0.33$


\textit{(ii)} PAH band strength ratios commonly used as proxies for the charge state of the PAH population are successfully calibrated against the PAH ionization parameter which allows for an estimate of the average ionization parameter in galaxies directly from observations. Furthermore, the correlation for the 6.2/11.2~\micron\ PAH band strength ratio and $\gamma$ naturally extends that found for the RN NGC~7023 to lower $\gamma$.

\textit{(iii)} A moderate correlation is found between the 8.6/11.2~\micron\ PAH ratio and sSFR, indicating an increase of PAH ionization in systems of younger stellar populations and recent star-formation episode, and a weak anti-correlation is observed for the PAH ionization parameter ($\equiv\gamma$) and M$_{*}$, which suggests a higher ionization efficiency for PAHs in low-mass galaxies. This could point to a slightly different make up of the PAH population, i.e., formation and evolution path ways, depending on M$_{\rm *}$. 

\textit{(iv)} The 3.3/11.2~\micron\ PAH band strength ratio, commonly used as a proxy for PAH size, is successfully calibrated against \aNc. Here, the 3.3~\micron\ PAH band strength is determined from extrapolating the PAHdb-fits to cover the bands.

\textit{(v)} A library of PAH emission spectral templates parametrized on average excitation energy, \aNc\ and $\overline{f_{i}}$ are provided. 


\begin{acknowledgments}

  We would like to thank the referee for the constructive comments and suggestions that have improved the clarity of this paper. A.M.'s research was supported by an appointment to the NASA Postdoctoral Program at NASA Ames Research Center, administered by the Oak Ridge Associated Universities through a contract with NASA. C.B. is grateful for an appointment at NASA Ames Research Center through the San Jos\'e State University Research Foundation (NNX17AJ88A). C.B., J.D.B. and L.J.A. acknowledge support from the Internal Scientist Funding Model (ISFM) Directed Work Package at NASA Ames titled: ``Laboratory Astrophysics -- The NASA Ames PAH IR Spectroscopic Database''. Usage of the Metropolis HPC Facility at the Crete Center for Quantum Complexity and Nanotechnology of the University of Crete, supported by the European Union Seventh Framework Programme (FP7- REGPOT-2012-2013-1) under grant agreement no. 316165, is also acknowledged. 

\end{acknowledgments}

\bibliography{bibliography}
\bibliographystyle{aasjournal}

\appendix

\section{Examples of decomposition and modeling for galaxy spectra of various properties} \label{sec:additional_examples}

\begin{figure*}
    \includegraphics[width=\linewidth]{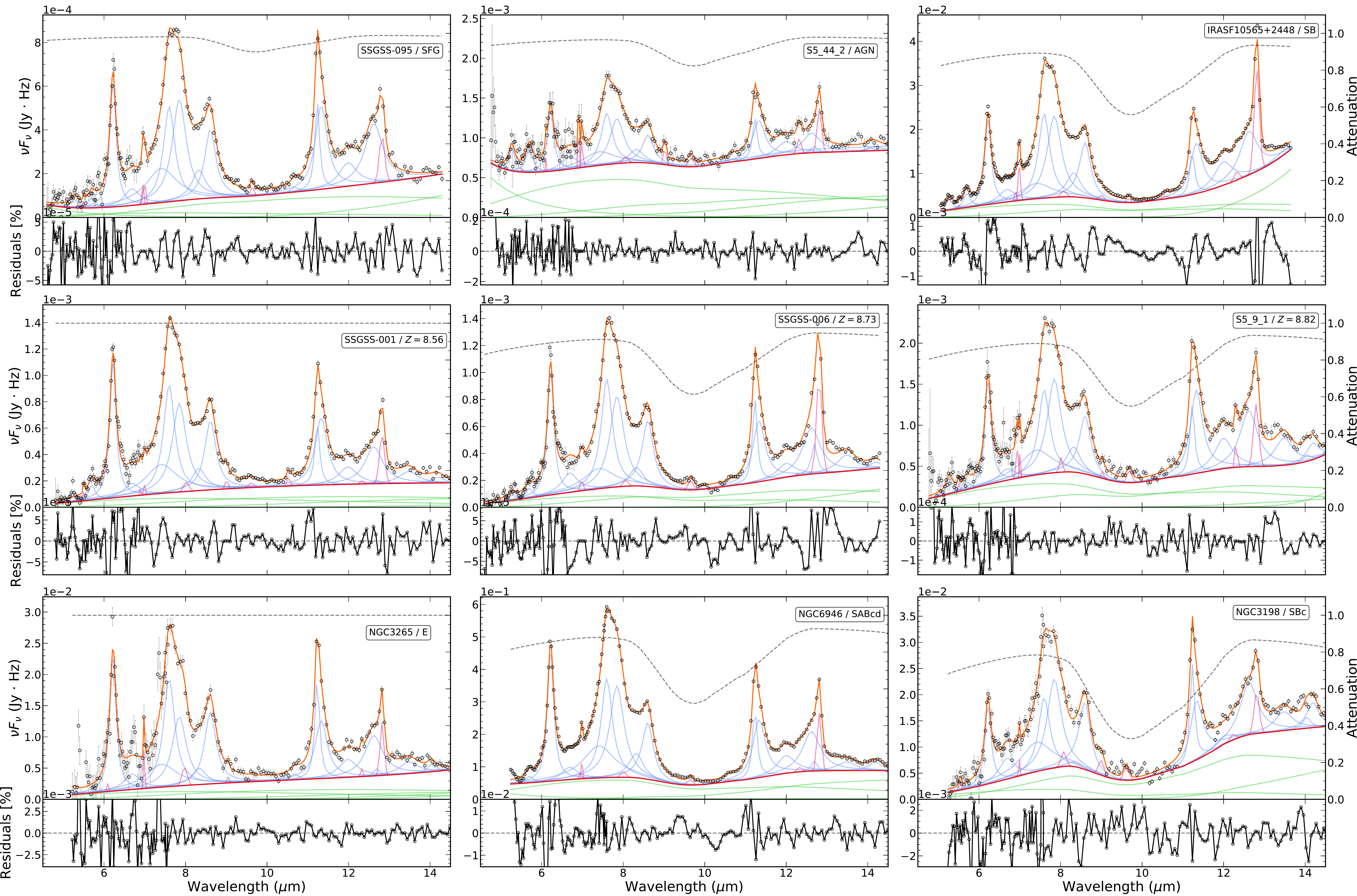}
        \caption{\textsc{pahfit} decomposition of spectra from different galaxy classes and properties. Top row: A main sequence SFG (left), an AGN (middle), and a starburst galaxy (right). Middle row: Spectra at different metallicities; $Z=8.56$ (left), $Z=8.73$ (middle), and $Z=8.82$ (right). Bottom row: Different morphological types; Elliptical (left), SABcd (middle), and SBc (right). See Figure~\ref{fig:pahfit} for the color coding.}
    \label{fig:pahfit_examples}
\end{figure*}

\begin{figure*}
    \includegraphics[width=\linewidth]{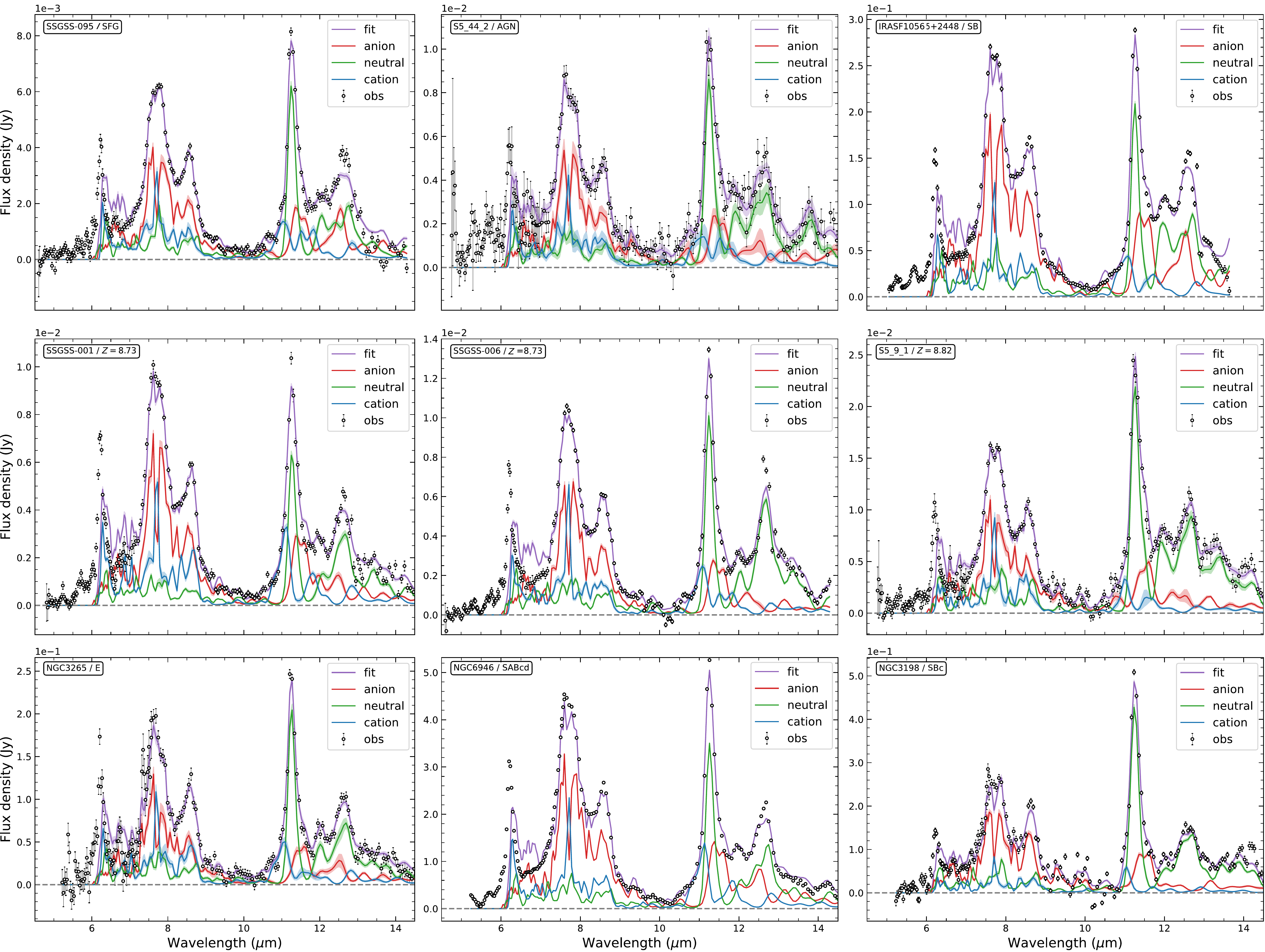}
    \caption{PAHdb charge breakdown of spectra from different galaxy classes and properties. Top row: A main sequence SFG (left), an AGN (middle), and a starburst galaxy (right). Middle row: Spectra at different metallicities; $Z=8.56$ (left), $Z=8.73$ (middle), and $Z=8.82$ (right). Bottom row: Different morphological types; Elliptical (left), SABcd (middle), and SBc (right).}
    \label{fig:pahdb_examples}
\end{figure*}

Figure \ref{fig:pahfit_examples} presents \textsc{pahfit} decomposition examples of spectra from different galaxy classes (SFG, AGN, starburst) and properties (metallicity, morphological type). For the different classes, a main sequence SFG, and AGN verified in both optical (BPT) and EQW$_{\rm 6.2}$ activity diagnostics, and a starburst of SFR $>100$ M$_{\odot}/$yr, were selected. Metallicity examples were drawn from galaxies with metallicity estimates of similar methods (i.e., MPA-JHU catalog). Examples of different galaxy morphology were drawn from the SINGS sample with available morphological classifications. Similarly, PAHdb charge breakdown for the same galaxies, based on their MC sampling modeling, are presented in Figure~\ref{fig:pahdb_examples}.

\section{Relationship between PAH band strength ratios with galaxy and PAHdb properties} \label{sec:pahratio_pairplots}

\begin{figure*}
    \centering
    \includegraphics[width=\linewidth]{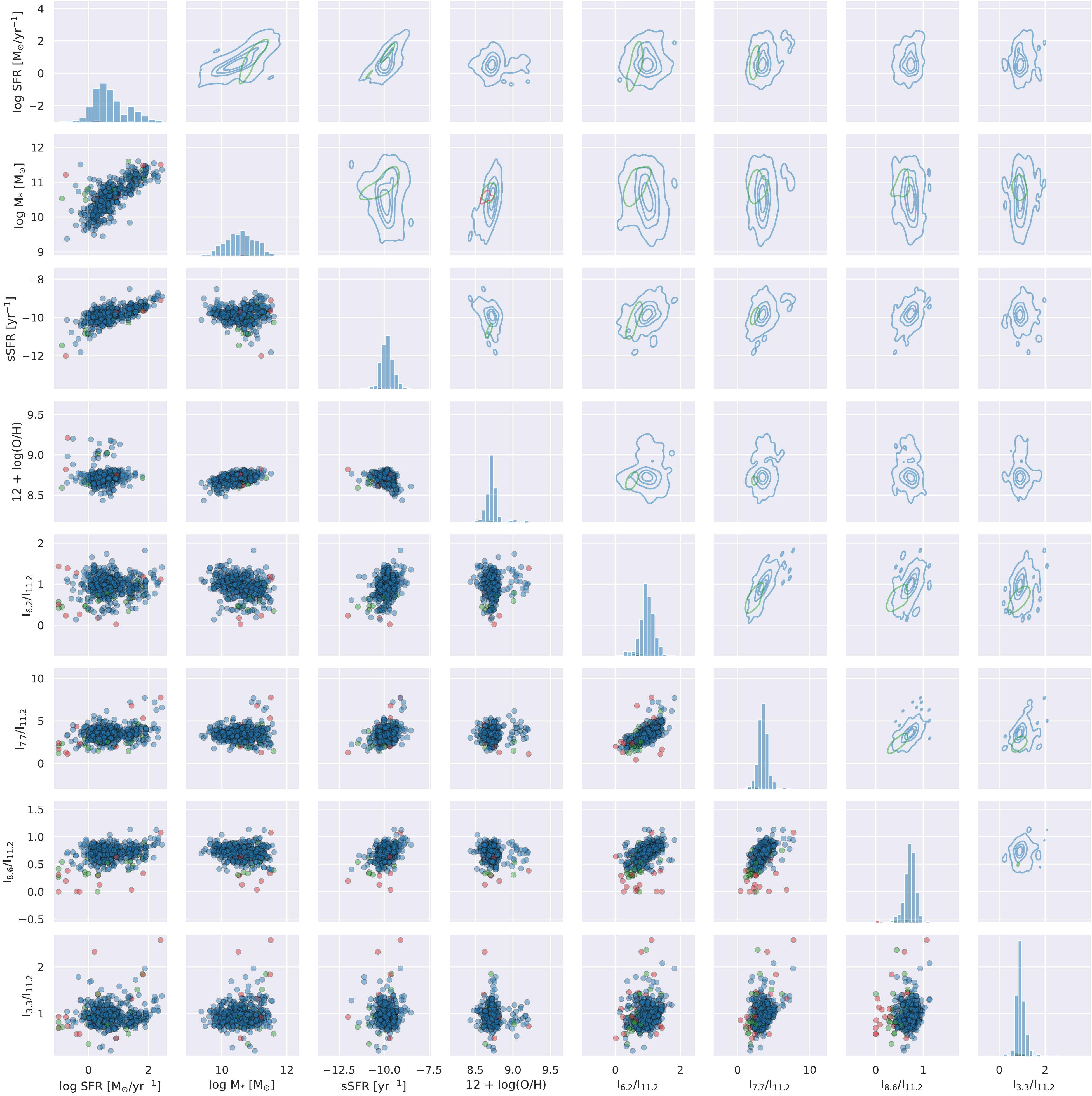}
    \caption{Pairwise relationships of PAH band strength ratios (I$_{6.2}$/I$_{11.2}$, I$_{7.7}$/I$_{11.2}$, I$_{8.6}$/I$_{11.2}$, I$_{3.3}$/I$_{11.2}$) and galaxy (SFR, M$_{*}$, sSFR, 12+log(O/H)) properties (points), along with their respective distributions (histograms), and kernel density estimations (contours at 5 levels, corresponding to iso-proportions of the probability mass density, i.e., contours at 80, 60, 40, and 20\%). Colors correspond to galaxy activity classes, with SFGs shown in blue, AGN in red, and CO in green. \label{fig:PAH_ratios-galaxy_pairplots}}
\end{figure*}

\begin{figure*}
    \centering
    \includegraphics[width=\linewidth]{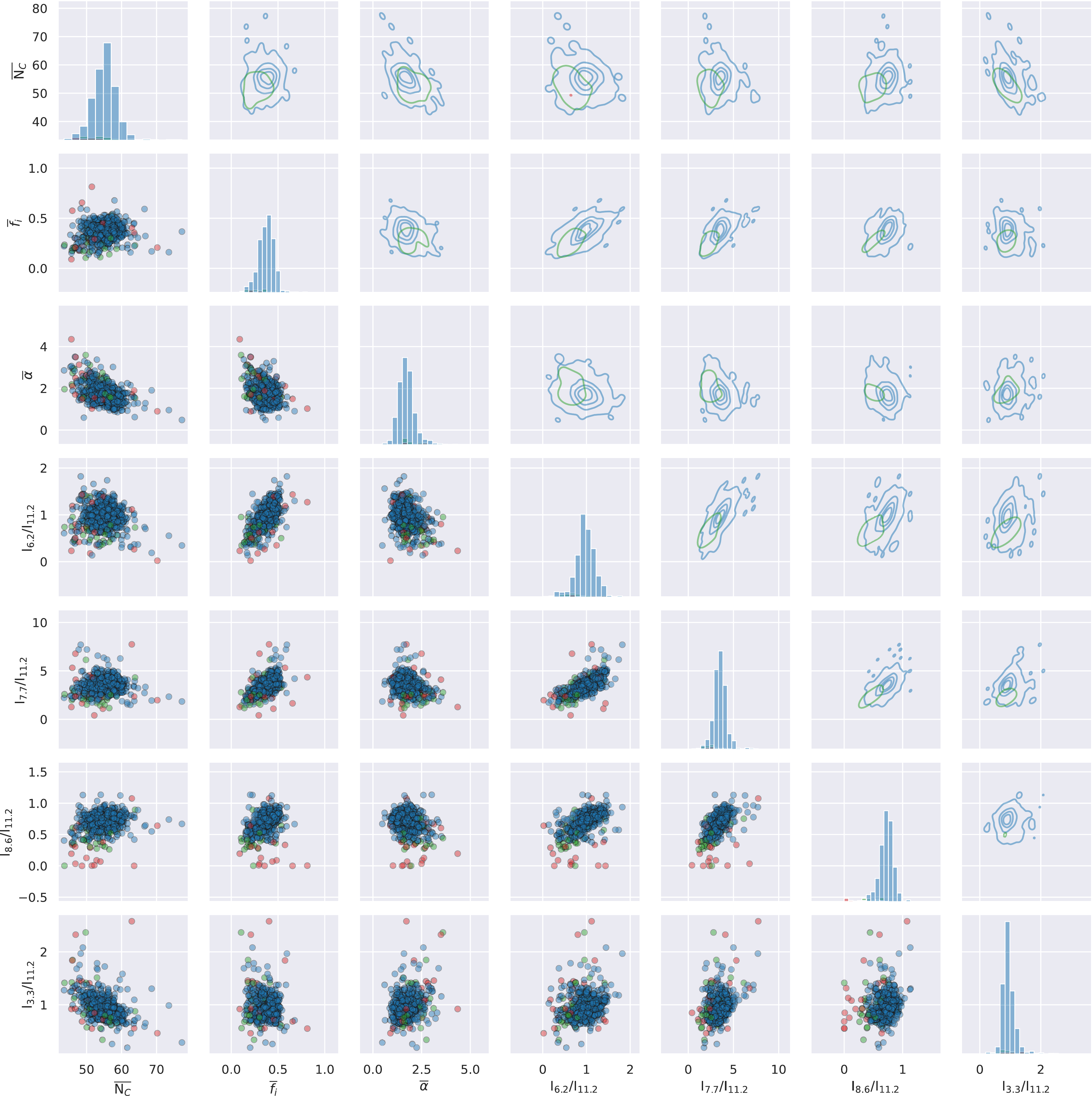}
    \caption{Pairwise relationships of PAH band strength ratios (I$_{6.2}$/I$_{11.2}$, I$_{7.7}$/I$_{11.2}$, I$_{8.6}$/I$_{11.2}$, I$_{3.3}$/I$_{11.2}$) and PAH (\aNc, \afi, $\overline{\alpha}$) properties (points), along with their respective distributions (histograms), and kernel density estimations (contours at 5 levels, corresponding to iso-proportions of the probability mass density, i.e., contours at 80, 60, 40, and 20\%). Colors correspond to galaxy activity classes, with SFGs shown in blue, AGN in red, and CO in green. \label{fig:PAH_ratios-PAHdb_pairplots}}
\end{figure*}

Figures \ref{fig:PAH_ratios-galaxy_pairplots} and \ref{fig:PAH_ratios-PAHdb_pairplots} present the pairwise relationship between PAH band strength ratios (I$_{6.2}$/I$_{11.2}$, I$_{7.7}$/I$_{11.2}$, I$_{8.6}$/I$_{11.2}$, I$_{3.3}$/I$_{11.2}$) with galaxy properties (SFR, M$_{*}$, sSFR, 12+log(O/H)) and PAHdb-derived properties (\aNc, \afi, $\overline{\alpha}$) respectively.

\section{Weak or Absent Correlations} \label{sec:weaker}

\begin{figure*}
    \includegraphics[width=\linewidth]{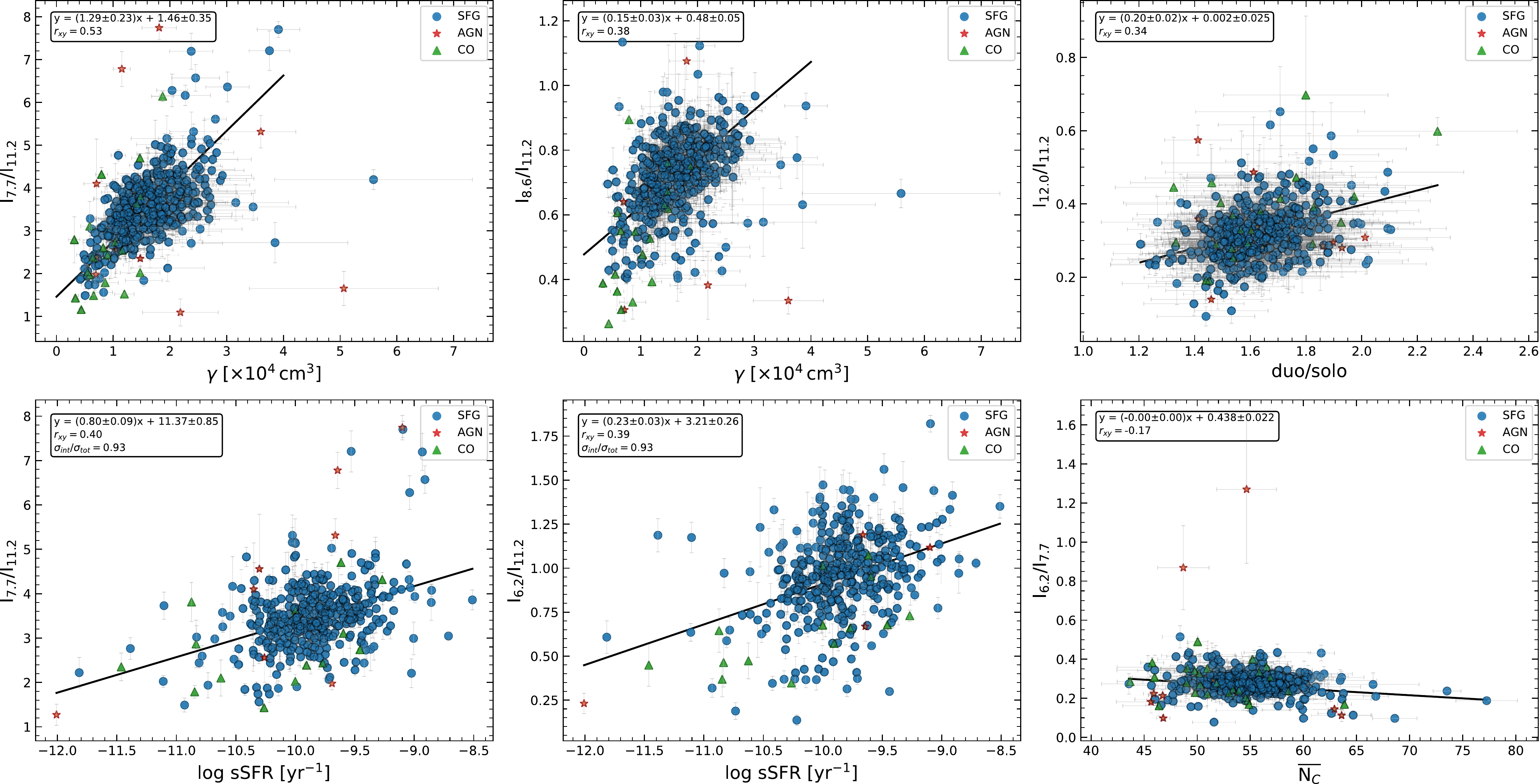}
    \caption{Relationship between PAH band strength ratios and PAHdb-dreived parameters showing weaker or no correlations. \textbf{Top left:} I$_{7.7}$/I$_{11.2}$ vs $\gamma$; Top middle: I$_{8.6}$/I$_{11.2}$ vs $\gamma$; \textbf{Top right:} I$_{12.0}$/I$_{11.2}$ vs duo/solo; \textbf{Bottom left:} I$_{7.7}$/I$_{11.2}$ vs sSFR; \textbf{Bottom middle:} I$_{6.2}$/I$_{11.2}$ vs sSFR; \textbf{Bottom right:} I$_{6.2}$/I$_{7.7}$ vs \aNc. The fit equations are given in the box, together with the Pearson's correlation coefficient ($r_{xy}$), and the ratio of the intrinsic to the total scatter ($\sigma_{int}/\sigma_{tot}$) where present.} 
    \label{fig:secondary}
\end{figure*}

Figure~\ref{fig:secondary} presents the relationship between PAH band strength ratios and PAHdb-derived parameters with weak or absent correlations. See also the discussion in Section~\ref{subsec:pahratios}. 

\section{Comparison between Quality Classes} \label{sec:classcomp}

\begin{figure*}
    \includegraphics[width=\textwidth]{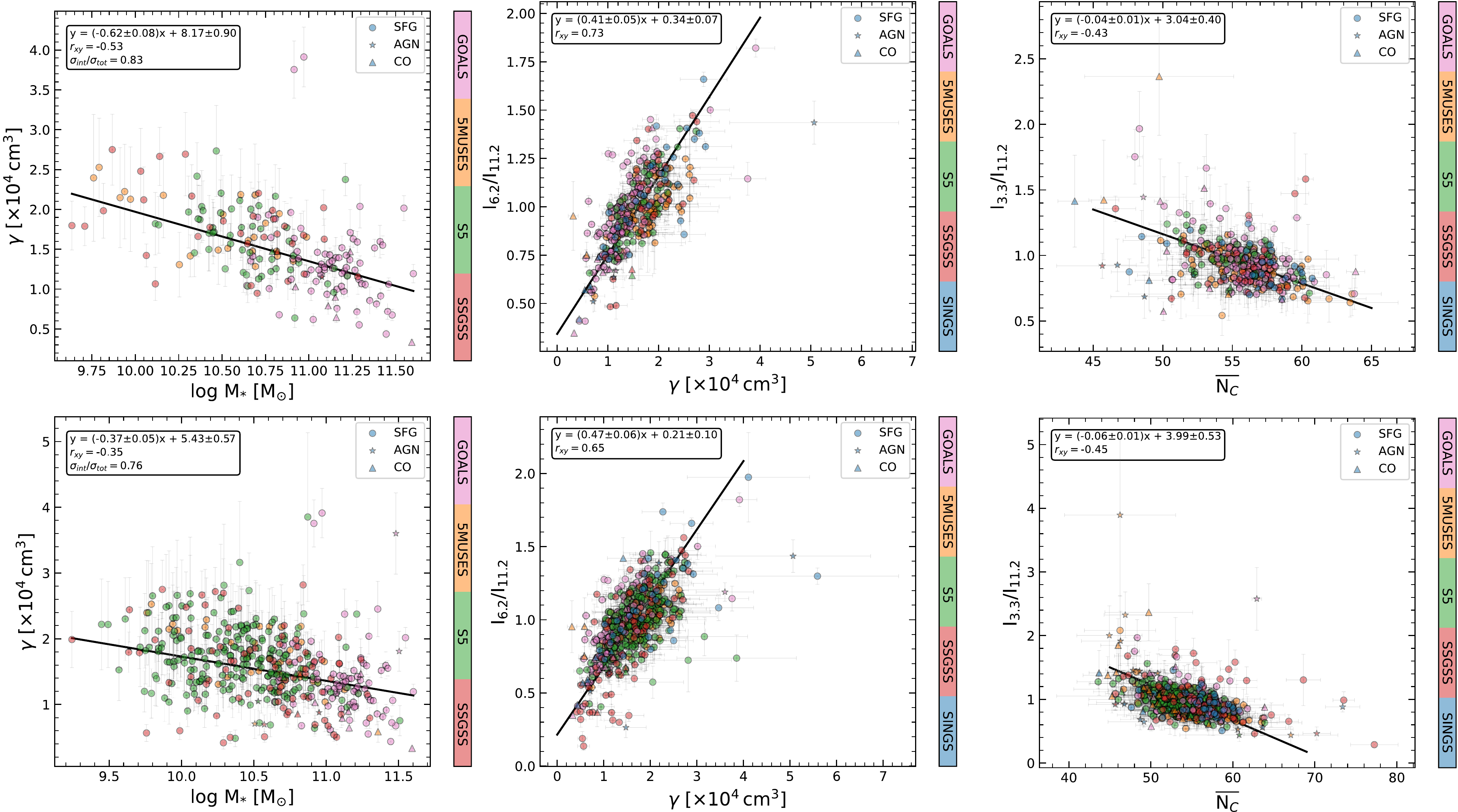}
    \caption{Correlations for $\gamma$ and stellar mass (left panels), the 6.2/11.2~\micron\ PAH band strength ratio and $\gamma$ (middle panels), and the 3.3/11.2~\micron\ PAH band strength ratio and \aNc\ (right panels) for the Q1 (top row) and Q3 (bottom row) quality classes. SFGs are shown as circles, AGN as stars, and CO as triangles. Data points are color-coded based on their associated Legacy program. The linear fit is shown as the black line and the fit equation is given in the legend, together with the Pearson's correlation coefficient ($r_{xy}$), and the ratio of the intrinsic to the total scatter ($\sigma_{int}/\sigma_{tot}$) where present.}
    \label{fig:Q1Q3}
\end{figure*}

Three quality classes were defined (Q1-Q3; cf.\ Section~\ref{sec:Results}, with Q1 being the most restrictive. Figure~\ref{fig:Q1Q3} presents the correlation between $\gamma$ and M$_{\rm *}$, the 6.2/11.2~\micron\ PAH band strength ratio with $\gamma$, and the 3.3/11.2~\micron\ PAH band strength ratio with \aNc\ for Q1 and Q3. Those for Q2 are presented in Figures~\ref{fig:PAH_gal} and \ref{fig:pahint}. The anti-correlation between $\gamma$ and M$_{*}$ improves when going to Q1 ($r_{xy}$=-0.53). Similarly, there is a tighter correlation between the 6.2/11.2~\micron\ PAH band strength ratio and $\gamma$ for Q1 ($r_{xy}$=0.73). The correlation for the 3.3/11.2~\micron\ PAH band strength ratio and \aNc\ is well established and shows no real differences between the different quality classes.

\section{Types of PAH Spectra} \label{sec:spectypes}
\subsection{Observed vs Modeled Isolated PAH Spectrum} \label{subsec:obsvsmod}

\begin{figure*}

    \includegraphics[width=\textwidth]{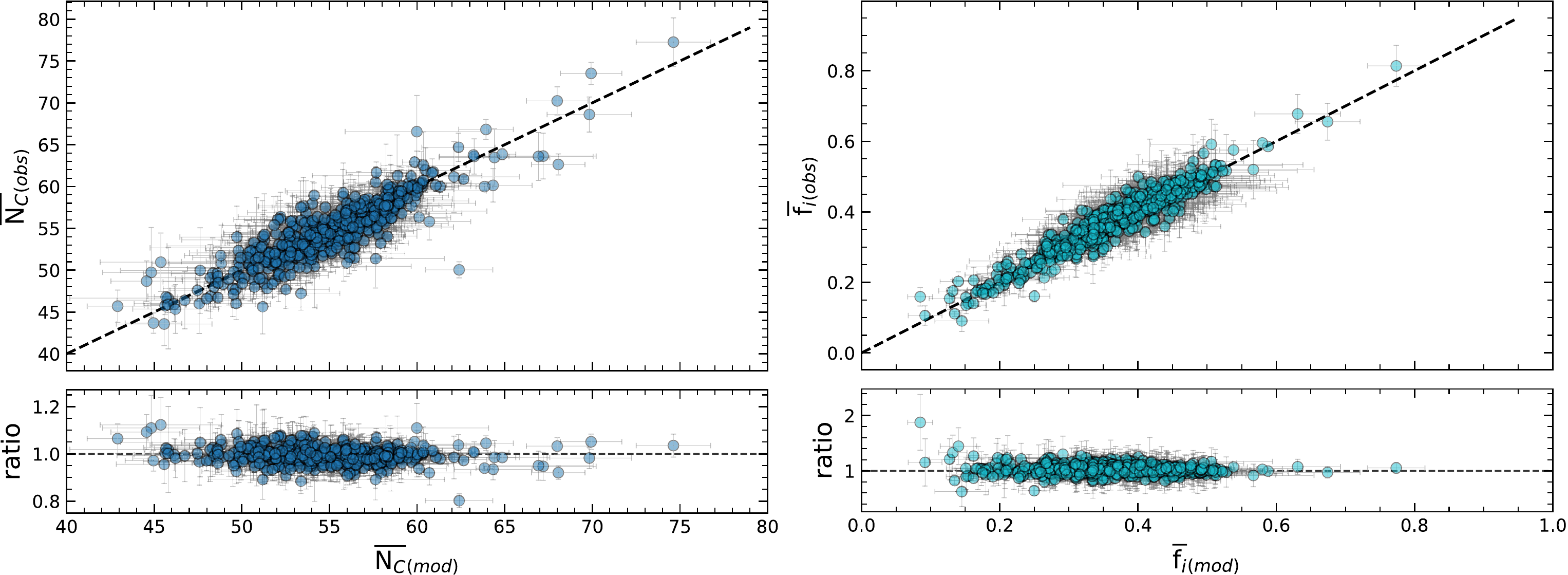}
    \caption{Comparison between \aNc\ (left panel) and $\overline{f_{i}}$ (right panel) derived from  the observed and modeled isolated Q2 PAH spectrum. The dashed line is the line of equality. In the bottom panels  the two derived quantities are ratio-ed.}
    \label{fig:PAHdb_obs-mod}
\end{figure*}

Two types of isolated PAH spectra have been considered: \textit{(i)} The \textsc{pahfit} modeled PAH spectrum, and \textit{(ii)} that constructed by subtracting the \textsc{pahfit} dust continuum, stellar continuum, atomic and H$_{2}$ line components, from the observed spectrum, and accounting for extinction. Figure~\ref{fig:PAHdb_obs-mod} compares \aNc\ and $\overline{f_{c}}$ for the two spectral types.

The figure shows that using either type of isolated PAH spectrum gives, within the uncertainties, consistent results. Each method has its benefit. For example, the isolated observed PAH spectrum retains residual PAH emission that is not matched by any of the \textsc{pahfit} components. On the other hand for low S/N spectra, the modeled PAH spectrum will be able to reconstruct some of band structure of individual features \citep[][]{Boersma2016}.

\subsection{Radiation Fields} \label{subsec:rfields}

\begin{figure*}
    \includegraphics[width=\linewidth]{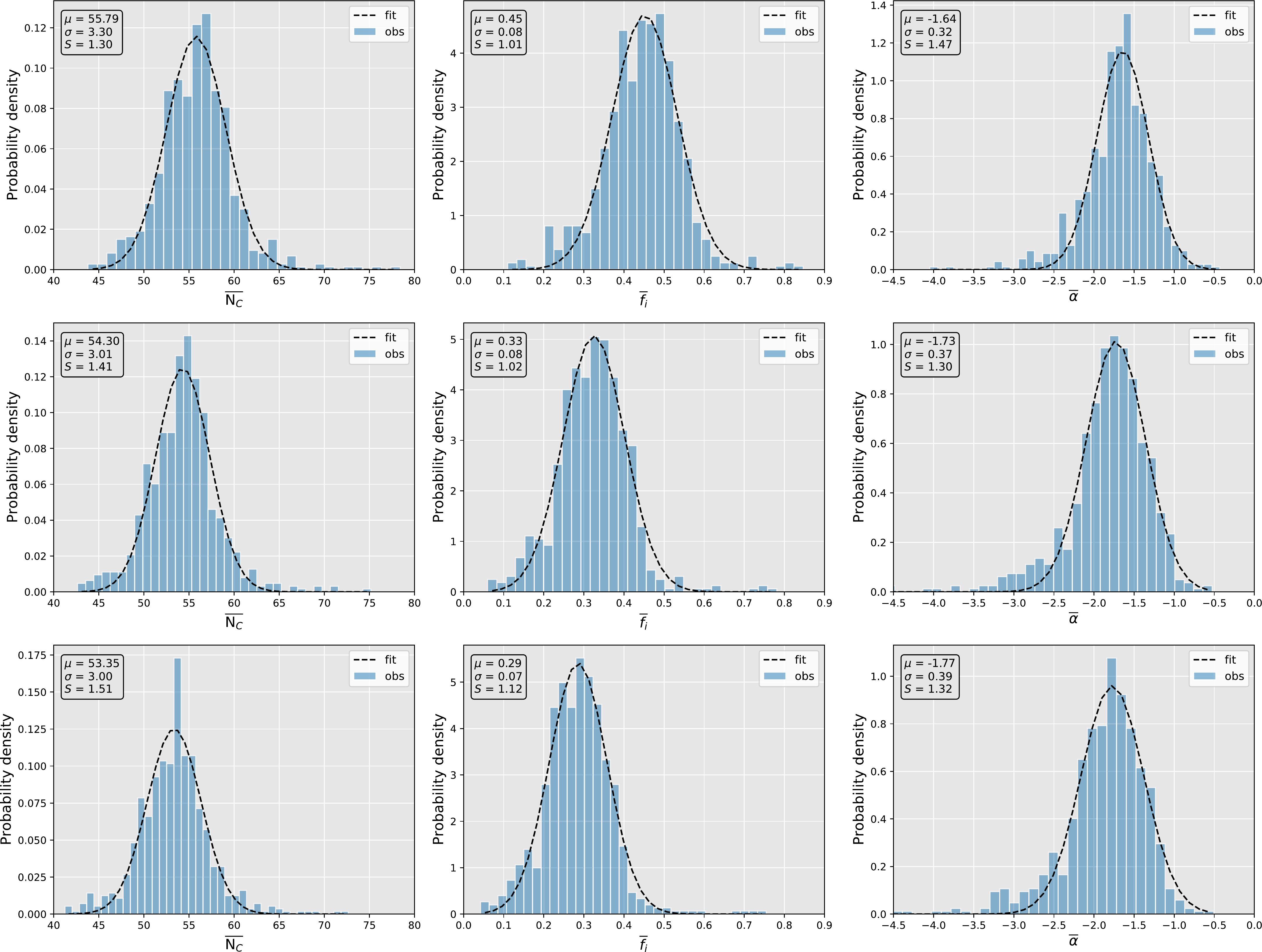}
    \caption{Distributions (Q2) for \aNc\ (left panels), \afi\ (middle panels), and $\overline{\alpha}$ (right panels) at an excitation energy of 6~eV (top row), 10~eV (middle row) and 12~eV. Distributions are fitted with a Gaussian model (black dashed lines). The statistical mean ($\mu$), standard deviation ($\sigma$), and skewness (\textit{S}) are given in the boxes.}
    \label{fig:radfieldcomp}
\end{figure*}

Figure~\ref{fig:radfieldcomp} presents the distribution of the PAHdb derived properties \aNc, $\overline{f_{i}}$, and $\overline{\alpha}$ for the three other excitation energies considered; 6, 10, and 12~eV. \aNc\ and $\overline{f{i}}$ show a small shift towards lower values with increasing photon energy but remain consistent within their uncertainties.

\section{The PAHdb Fitted 6.2 \texorpdfstring{\micron}{um} PAH band} \label{sec:PAHdbPAH62}

\begin{figure*}
    \includegraphics[width=\textwidth]{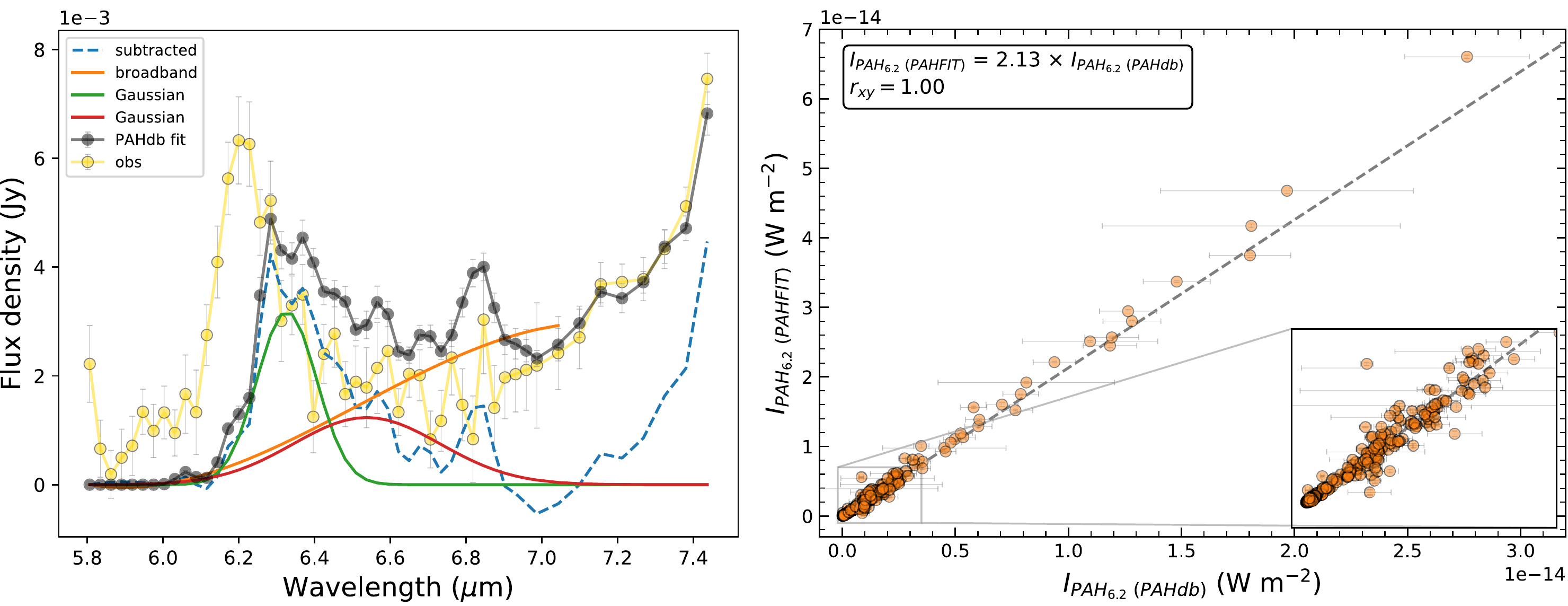}
    \caption{\textbf{Left:} Demonstration of the approach used for determining the amount of the 6.2~\micron\ PAH band matched by PAHdb for the spectrum (black line and points) of the galaxy SDSS~J093001.33+390242. First, a spline (orange line) is fitted to match the blue wing of the broad dust feature component at 7.42~\micron, as employed by \textsc{pahfit}, and then is subtracted. The resulting subtracted spectrum (blue dashed line), is fitted with two Gaussians corresponding to the 6.2 and 6.69~\micron\ components used in \textsc{pahfit} (green and red lines). The observed PAH spectrum is shown for comparison (yellow line and points). \textbf{Right:} Comparison of the strength of the 6.2~\micron\ PAH band matched by \textsc{pahfit} and PAHdb. The gray dashed line shows a straight line fit (slope $=2.13$). }
    \label{fig:PAHdb_PAH62}
\end{figure*}

Here, we compare the 6.2~\micron\ PAH band strength matched by PAHdb and \textsc{pahfit}. We use the Q2 spectra with a S/N of at least 3 for the 6.2~\micron\ PAH band. To determine the match by PAHdb, we take the average 5.8-7.4~\micron\ fitted spectra from the MC sampling and: \textit{(i)} fit a spline through fixed anchor points to model the blue wing of the broad dust feature component used by \textsc{pahfit} at 7.42~\micron\ extending to the 6.2~\micron\ region; \textit{(ii)} use two Gaussians to fit the 6.2~\micron\ PAH band and the smaller blended dust feature component at 6.69~\micron\ as used by \textsc{pahfit}. Figure~\ref{fig:PAHdb_PAH62} (left panel) demonstrates this approach for the average MC PAHdb-fitted spectrum of the galaxy SDSS~J093001.33+390242. The right panel of the same figure compares the 6.2~\micron PAH band strength determined from the fitted Gaussians and that by \textsc{pahfit}. Fitting a straight line returns a slope of 2.13, while the Pearson's coefficient is r$_{xy}=1.00$.

\section{Available Online Data} \label{sec:mastertable}

The collected and derived galaxy properties, spectral quality parameters, \textsc{pahfit}-derived PAH band strengths and parameters, as well as PAHdb-derived parameters have all been made available online\footnote{\href{https://www.astrochemistry.org/pah\_galaxy\_properties/}{www.astrochemistry.org/pah\_galaxy\_properties/}}. Table~\ref{tab:machine} provides an overview of the available data with a short description.

\begin{deluxetable*}{ll|ll|ll}
  \tablenum{4}
  \tabletypesize{\small}
  \tablecaption{Available Online Data \label{tab:machine}}
  \tablehead{\multicolumn{2}{c}{Galaxy Properties} & \multicolumn{2}{c}{\textsc{pahfit}-Derived Properties} & \multicolumn{2}{c}{PAHdb-Derived Properties} \\
  \colhead{Column} & \colhead{Description} & \colhead{Column} & \colhead{Description} & \colhead{Column} & \colhead{Description}
  }
  \startdata
  Galaxy & Galaxy name & Galaxy & Galaxy name & Galaxy & Galaxy name \\
  Sample & Legacy program & Sample & Legacy program & Sample & Legacy program \\
  Class & BPT classification & I$_{\rm 6.2}$ & 6.2~\micron\ PAH band strength & \aNc & Average number of carbon atoms\\
  EQW Class & EQW$_{\rm 6.2}$ classification & I$_{\rm 6.2\, unc}$ & 6.2~\micron\ PAH band strength unc & \aNc{} std & Number of carbon atoms std\\
  Dist & Distance & I$_{\rm 6.2\, EQW}$ & 6.2~\micron\ PAH band EQW & Cation & Cation fraction\\
  $z$ & Redshift & I$_{\rm 6.2\, EQW\,unc}$ & 6.2~\micron\ PAH band EQW unc & Cation std & Cation fraction std\\
  SFR & Star Formation Rate & I$_{\rm 7.7}$ & 7.7~\micron\ PAH band strength & Small & Small PAHs fraction\\
  M$_{*}$ & Stellar Mass & I$_{\rm 7.7\,unc}$ & 7.7~\micron\ PAH band strength unc & Small std & Small PAHs fraction std\\
  sSFR & Specific SFR & I$_{\rm 7.7\,EQW}$ & 7.7~\micron\ PAH band EQW & Large & Large PAHs fraction \\
  Z & Metallicity & I$_{\rm 7.7\,EQW\,unc}$ & 7.7~\micron\ PAH band EQW unc & Large std & Large PAHs fraction std \\
  SNR & S/N$_{\textrm{(tot)}}$ & I$_{\rm 8.6}$ & 8.6~\micron\ PAH band strength & Nitrogen & Nitrogen containing PAH fraction\\
  SNR 11.2 & S/N$_{(11.2)}$ & I$_{\rm 8.6\,unc}$ & 8.6~\micron\ PAH band strength unc & Nitrogen std & Nitrogen containing PAH fraction std\\
  Q & Quality Class & I$_{\rm 8.6\,EQW}$ & 8.6~\micron\ PAH band EQW & Pure & Pure PAHs fraction\\
  & & I$_{\rm 8.6\,EQW\,unc}$ & 8.6~\micron\ PAH band EQW unc & Pure std & Pure PAHs fraction std\\
  & & I$_{\rm 11.2}$ & 11.2~\micron\ PAH band strength & Ion frac & Ionization fraction\\
  & & I$_{\rm 11.2\,unc}$ & 11.2~\micron\ PAH band strength unc & Ion frac std & Ionization fraction std\\
  & & I$_{\rm 11.2\,EQW}$ & 11.2~\micron\ PAH band EQW & Ioniz param & Ionization parameter\\
  & & I$_{\rm 11.2\,EQW\,unc}$ & 11.2~\micron\ PAH band EQW unc & Ioniz param unc & Ionization parameter unc\\
  & & I$_{\rm 12.0}$ & 12.0~\micron\ PAH band strength & a$_{eff}$ & PAH effective radius\\
  & & I$_{\rm 12.0\,unc}$ & 12.0~\micron\ PAH band strength unc & a$_{eff}$ std & PAH effective radius std\\
  & & I$_{\rm 12.0\,EQW}$ & 12.0~\micron\ PAH band EQW & $\alpha$ & a$_{eff}$ distribution power-law index\\
  & & I$_{\rm 12.0\,EQW\,unc}$ & 12.0~\micron\ PAH band EQW unc & $\alpha$ std & a$_{eff}$ distribution power-law index std\\
  & & S07 att & $\tau_{9.7}$ silicate & avg solo & average number of solo hydrogens\\
  & & S07 att unc & $\tau_{9.7}$ silicate unc & avg solo std & number of solo hydrogens std\\
  & & $\sigma_{\textsc{pahfit}}$ & \textsc{pahfit} unc & avg duo & average number of duo hydrogens\\
  & & Q & Quality Class & avg duo std & number of duo hydrogens std\\
  & & & & avg trio & number of trio hydrogens \\
  & & & & avg trio std & number of trio hydrogens std\\
  & & & & avg quartet & number of quartet hydrogens \\
  & & & & avg quartet std & number of quartet hydrogens std\\
  & & & & I$_{\rm 3.3}$ & 3.3~\micron\ PAH band strength\\
  & & & & I$_{\rm 3.3\,unc}$ & 3.3~\micron\ PAH band strength unc\\
  & & & & $\sigma_{PAHdb}$ & average PAHdb unc\\
  & & & & $\sigma_{PAHdb}$ std & PAHdb unc std\\
  & & & & $\sigma_{PAHdb\,11.2}$ & average PAHdb 11.2~\micron\ unc\\
  & & & & $\sigma_{PAHdb\,11.2}$ std & PAHdb 11.2~\micron\ unc std\\
  \enddata
\end{deluxetable*}

\section{Comparison between IDL and Python PAHFIT} \label{sec:IDLvsPy}

\begin{figure*}
    \includegraphics[width=\linewidth]{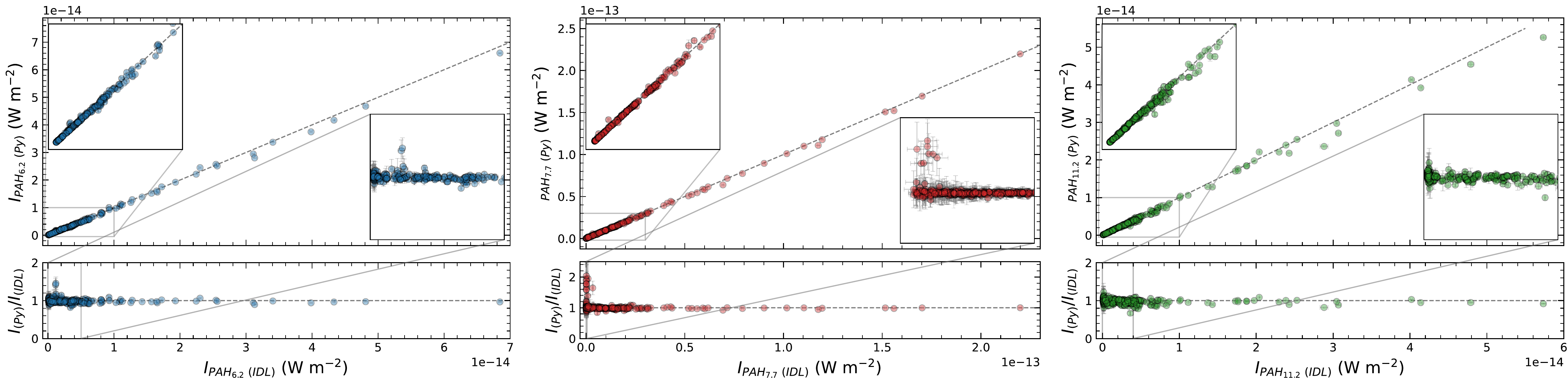}
    \caption{Comparison between the Q2 6.2, 7.7, and 11.2~\micron\ PAH band strengths determined using the IDL and Python version of \textsc{pahfit}. Lines of equality are drawn in the upper panels (gray dashed). The bottom panels show the ratio of the Python over IDL \textsc{pahfit}-determined PAH band strengths.}
    \label{fig:IDLvsPy}
\end{figure*}

Figure~\ref{fig:IDLvsPy} compares the Q2 6.2, 7.7, and 11.2 \micron\ PAH band strengths determined using the IDL and Python version of \textsc{pahfit}. The figure shows good agreement, with most of the scatter for sources with low PAH band strengths. Only galaxies with PAH band strengths greater than 10$^{\rm -16}$ W m$^{\rm -2}$ and a S/N of at least 3 are considered.

\end{document}